\documentclass[10pt,journal]{IEEEtran}

\ifCLASSINFOpdf

\else

\fi

\usepackage{amsmath}
\usepackage{amsthm}
\newtheorem{Definition}{Definition}
\newtheorem{Theorem}{Theorem}
\newtheorem*{Proof}{Proof}

\usepackage{svg}
\usepackage{subfigure}
\usepackage{amssymb}
\usepackage{booktabs}
\usepackage{multirow}
\begin{document}

\title{Protection Degree and Migration in the Stochastic SIRS Model: A Queueing System Perspective}

\author{Yuhan Li, Ziyan Zeng, Minyu Feng {\em Member, IEEE}, and J\"{u}rgen Kurths

\thanks{Manuscript received August 11, 2021; revised September 25, 2021 and October 5, 2021; accepted October 6, 2021. Date of publication October 19, 2021; date of current version January 28, 2022. This work was supported
in part by the Ministry of Education in China (MOE) Project of Humanities and Social Sciences under Grant 21YJCZH028, in part by the Fundamental Research Funds for the Central Universities under Grant SWU019029, and in part by the Ministry of Science and Higher Education of the Russian Federation within the framework of state support for the creation and development of World-Class Research Centers “Digital Biodesign and Personalized Healthcare” under Grant 075-15-2020-926. This article was recommended by Associate Editor E. Tlelo-Cuautle. (Corresponding author: Minyu Feng.)}

\thanks{Yuhan Li, Ziyan Zeng, Minyu Feng (corresponding author) are with College
of Artificial Intelligence, Southwest University, Chongqing 400715, P. R. China.
(e-mail: myfeng@swu.edu.cn).
}

\thanks{J\"{u}rgen Kurths is with Potsdam Institute for Climate Impact Research, 14437 Potsdam, Germany,
and also with the Centre for Analysis of Complex Systems, World-Class Research Center "Digital biodesign and personalized healthcare", Sechenov First Moscow State Medical University, Moscow, 119991Russia.}

\thanks{Color versions of one or more figures in this article are available at
https://doi.org/10.1109/TCSI.2021.3119978.}

\thanks{Digital Object Identifier 10.1109/TCSI.2021.3119978}}

\mark{IEEE TRANSACTIONS ON CIRCUITS AND SYSTEMS I: REGULAR PAPERS. }%

\maketitle

\begin{abstract}
With the prevalence of COVID-19, the modeling of epidemic propagation and its analyses have played a significant role in controlling epidemics. However, individual behaviors, in particular the self-protection and migration, which have a strong influence on epidemic propagation, were always neglected in previous studies. In this paper, we mainly propose two models from the individual and population perspectives. In the first individual model, we introduce the individual protection degree that effectively suppresses the epidemic level as a stochastic variable to the SIRS model. In the alternative population model, an open Markov queueing network is constructed to investigate the individual number of each epidemic state, and we present an evolving population network via the migration of people. Besides, stochastic methods are applied to analyze both models. In various simulations, the infected probability, the number of individuals in each state and its limited distribution are demonstrated.
\end{abstract}


%
\IEEEpeerreviewmaketitle

\section{Introduction}
As is well known, epidemics are the enemy faced by all of mankind and pose an enormous threat to human health and society. There are infectious diseases appearing and humans have been fighting against them all the time. From bubonic plague to SARS followed by COVID-19, there is a desire for controlling epidemics. Thanks to the increasing concern and awareness of diseases, people urgently want to understand the pathogenesis and propagation mechanism of epidemics. Consequently, a number of researchers began to study the propagation of epidemics increasingly in-depth and proposed various models expecting to provide some solutions to epidemic prevention and control.

Kermack and McKendrick first proposed the SIR compartment model \cite{SIR} and later the SIS model, which laid a foundation for studies on epidemics. In the light of complex networks, researchers began to combine networks with epidemic models. Based on the typical SIS and SIR model, as an improvement of the SIR epidemic model, the SEIR epidemic model which includes the incubation phase was proposed and studied on various networks, and different types of the epidemic model was studied combined with networks, e.g., layered networks \cite{layer}, small-world networks \cite{WS}, temporal networks \cite{tem}, and multiplex networks \cite{ieeesis}. Besides, the SAIR model was established to describe an asymptomatic phase as an extension of the SIR model \cite{SAIR}. With the development of the big data, utilizing data to track epidemics is useful for analyzing the spreading of epidemics \cite{wendong1}, \cite{wendong2}. Ref. \cite{ieeediff}, studied the diffusion of opinion and information, which was considered together with the epidemics spreading in some research. For nowadays COVID-19 pandemic, various models spring up, like the model with microscopic Markov-chain approach for studying spatiotemporal spreading of COVID-19 \cite{cov}, the social-network based model considering the heterogeneity of the population \cite{cov_revision}, and the model considering the travelling population and the lockdown strategy \cite{cov_re2}.

In addition to a variety of epidemic models considering different phases of epidemics, in recent years, individual behaviors and awareness were studied a lot in epidemic modeling. An SIS model regarding awareness weighted by both local information, i.e., the fraction of infected neighbors and global broadcast was proposed \cite{17aware}. The awareness leading to taking precautionary measures was considered and described via a reduction of the transmission probability by an exponential factor, which suppresses the prevalence of epidemics \cite{18aware}, and a general awareness-induced general and Brownian Motions was introduced in the SIRS model \cite{asmy}. In \cite{19aware}, an unaware-aware process was built with the SIR and evaluated the effect of the awareness with time step. A two-layered networks was established for analyzing multiple influences between awareness diffusion and epidemic propagation \cite{aware}. Coevolution of vaccination opinion and awareness was also investigated in a three-layered complex network \cite{ieeeco}. Later, in terms of the spatial-temporal properties, a layer-preference walk model based on the multiplex network which consists static information spreading network and a temporal physical contact network was studied \cite{21aware}. Besides, the individual behavior of wearing masks was considered in \cite{mask}, which evaluated the infection cost and the cost of wearing masks and discussed their effect on epidemics in different cases.

A lot of methods are applied to analyzing epidemic models, e.g., the classical and well-known mean-field method was utilized to study various dynamics systems\cite{meanfield}, \cite{ieeemeanfield}, and heterogeneous mean-field method (HMF) was proposed for the propagation on heterogenous networks \cite{hmf}. Later, the quench mean-field approach \cite{qmf}, Markov-chain approach \cite{MarkovG} and N-intertwined approach \cite{n} were established, which utilize the adjacent matrix of networks. Later, graph-coupled hidden Markov Models were proposed to study the spread on the individual level \cite{wendong}. With the exception of these typical methods, queueing theory was usefully applied to epidemic modeling in networks. Pieter Trapman et al. investigated the relation between the spread of epidemics and the dynamics of the queueing system which specifically is a M/G/1 queue, and focused on the infectious individual number in the SIR model at the moment of the first detection of the epidemic \cite{q0}. Queueing theory was also applied for SIS and SEIS epidemic models where the basic reproductive number was provided based on the queueing system approach \cite{q01}. In \cite{q02}, the typical SI epidemic process with a recovery rule was modeled as a queueing system for revealing its transient characteristics. Later, an epidemic Markov queueing model was proposed by constructing an M/M/1 queueing system with input and output flow transition rates \cite{q1}. A queueing-based compartmental model was developed to study the Ebola virus disease \cite{q11}. In \cite{q2}, researchers developed a novel metric of viral transmissibility in queueing systems with overlapping sojourn time.

Various epidemic models have been studied as described above, while there are still crucial issues on unfitting real situations, e.g., the lack of the impact of individuals awareness and migration on epidemic spreading. Though there are models considering the individual awareness, it is only described as a simple changeable parameter, ignoring its difference among individuals. Moreover, most epidemic spreading processes are studied on a static network. However, the population network of an area is always changing due to the complex reality factors which may cause the change of the network structure.

To improve the present epidemic models, in this paper, we propose a novel SIRS epidemic model considering the individual protection degree and the migration of mobile population for more appropriate modeling. We establish two Markov processes from two perspectives, respectively presenting the dynamics of transitions between individual states and the change of the individual number in three states. In the individual model, we reveal the impact of contacts and individual protective behaviors on propagation. In the population model, by virtue of queueing theory, we explain the mechanism of the change of the individual number via constructing a Markov queueing network. In the simulation, we demonstrate the impact of the protection degree on the epidemic, the number of individuals varying with time and the distribution of the individual number.

The organization of the paper is as follows: In Section \ref{sec:II}, we display the construction of proposed models and give analysis and theoretical results of the model. Simulations are carried out to demonstrate the validity of our models and theorems in Section \ref{sec:III}. Conclusions and future work are given in Section \ref{sec:IV}.

\section{\label{sec:II}EPIDEMIC SPREADING IN COMPLEX NETWORKS BASED ON MARKOV METHOD}
On the basis of typical epidemic models, we propose a new epidemic SIRS model regarding the migration of mobile population as well as the awareness of individual protection in an epidemic. We construct a network-based epidemic model. Suppose that each individual is a vertex in a network, and edges indicate contacts between vertices. Disease spreads throughout networks on account of infected individuals engaged in contacting the susceptible via links among them. And the network get evolved due to the migration of people. Additionally, the Markov method is applied to the SIRS epidemic modeling whose results can be displayed as a stationary value in a stochastic form distinguished from those converging to a specific value.

In terms of a living individual in our individual model, it randomly transforms among three states that are susceptible (S), infected (I) and recovered (R). Assume that the future state of an individual only depends on the current state and is independent of past states. The state of an individual can be therefore regarded as a non-homogeneous Markov chain, where three states are taken as the state space of Markov chain. We mark the three epidemic states with integers as 0, 1, 2. In the analysis below, we propose the transition matrix of states of individuals and display the iteration formulation of the limited distribution of three epidemic states.

In the population model, we take the number of individuals in different states separately as a continuous-time Markov process $\{N(t), t>0\}$ which takes on values in the set of nonnegative integers 0, 1, 2, $\cdots$. For each state, the number of the individual varies with time as a result of the migration of individuals as well as transitions between epidemic states. In this model, we focus on the expression of transition formulas and expectations of the individual number of three states. In this section, we model the epidemic on networks by constructing the above processes.
\subsection{\label{sec:II1}Markov Chain of Individual State}
\begin{figure}
\includegraphics[scale=0.9]{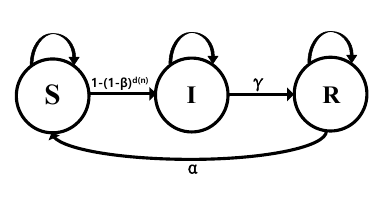}
\centering
\caption{State transition of SIRS model: The arrows indicate the transitions from one state to another, where the transition probabilities are given. The loop arrow indicates remaining staying in the present state without making a transition of an individual.}
\label{fig:SIRS}
\end{figure}
Primarily, we review the classical SIRS epidemic model. Let $S$ be the number of susceptible individuals, $N$ be the total number of individuals, and $\beta$, $\gamma$ be the given parameters. An infected individual has the ability to infect susceptible individuals at a rate ${\beta}S/N$ and recovers to become recovered at a constant rate $\gamma$ while recovered individuals avoid being infected but can be susceptible again by a certain probability $\alpha$ owing to the immune suppression.
Next, we creatively introduce the conception of protection degree of individuals to depict the behaviors that individuals improve the awareness of protecting themselves from infected people, which results in disconnections of contacts with infected ones. The reason for proposing the protection degree is that during an epidemic people are likely to cut down contacts to others to protect themselves. Thus, we take the protection degree into consideration to describe individual protective behaviors when modeling the epidemic propagation. We assume that the protection degree is a nonnegative variable following a Poisson distribution as definition \ref{d1}, which is reasonable for characterizing the randomness of the individual protection degree. The parameter of the Poisson distribution is also called the intensity typically denoted as $\lambda$, while we hereby denote by $\mu$ the parameter of the Poisson distribution to avoid the same notation of the input rate in the following section.
\begin{Definition}\label{d1}
Let $U$ denote the protection degree, which is a continuous random variable following the Poisson distribution, expressed as
\begin{equation}\label{eq:normal}
U\sim P(\mu).
\end{equation}
where $\mu$ is the parameter of the Poisson distribution.
\end{Definition}
In real situations, the higher degree of protection an individual leads to the higher probability that he breaks edges with infected ones, which leads to less infected neighbors in terms of the network. Therefore, we transform the protection degree into the form $e^{-u}$ with a range from 0 to 1. In this way, we define the valid rate of infected neighbors $f$,
\begin{equation}
f=e^{-u}.
\end{equation}
Suppose that $\rho_{j}(n)$ is the original number of infected neighbors of node $j$ at the time step $n$. Then the number of valid neighbors $d_{j}(n)$ is expressed as
\begin{equation}
d_{j}(n)=f\cdot \rho_{j}(n).
\end{equation}

Then, we apply the Markov Chain method to construct the propagation from the perspective of individuality. The state of each individual in a population then can be regarded as a Markov chain $\{X_{n}, n=0, 1, 2\}$, where the integers $0, 1, 2$ respectively represent the susceptible, infected and recovered state.
We define the probability that an infected individual makes a transition into a recovered state as the recovery probability, denoted as $\gamma$. The probability of transferring from recovered to susceptible is the revivification probability $\alpha$. Assume that there are all susceptible individuals except one individual being infected at the beginning of the epidemic and others are all susceptible. The initial state of the first infected individual is $\pi_{0}=(0, 1, 0)$, and the initial state of all susceptible individuals is $\pi_{0}=(1, 0, 0)$. In the transition process, for each node, the transition probability from the infected state to the recovered state is
$P_{1,2}=\gamma$,
and the transition probability from the recovered state to the susceptible one is
$P_{2,0}=\alpha$.
The crux is the infected probability related to the contacts to infected people. Namely, the probability that a susceptible individual transfers into the infected state is relevant to the degree and infected neighbor density of that node in the network. Then we have the transition probability of node $j$ transforming from susceptible to infected
\begin{equation}
P_{0,1}=1-(1-\beta)^{f{\cdot}\rho_{j}(n)}.
\end{equation}
Hence, denote by $d_{j}$ that the infected neighbor number at time step $n$ following $d_{j}(n)=k_{j}\rho_{j}(n)$. $\beta$ is infected rate as described above.

Having the above statements, we obtain the definition as follows.
\begin{Definition}\label{d2}
The transition matrix $P_{j}(n)$ of the susceptible individual $j$ at time step $n$ is
\begin{equation}\label{eq:4}
P_{j}(n)=\begin{bmatrix}
         (1-\beta)^{d_{j}(n)} & 1-(1-\beta)^{d_{j}(n)} & 0 \\
         0 & 1-\gamma & \gamma \\
         \alpha & 0 & 1-\alpha \\
       \end{bmatrix}.
\end{equation}
\end{Definition}

Based on Def. \ref{d2}, the epidemic state of an individual at $n+1$ can be obtained iteratively described as
\begin{equation}\label{it}
\pi_{n+1}=\pi_{0}P_{j}(0)P_{j}(1){\cdots}P_{j}(n)=\pi_{n}P_{n+1}.
\end{equation}

Additionally, through the transition matrix, we notice that $d_{j}(n)$ is a crucial element that affects the infected probability of a node. Intuitively, the number of infected neighbors of a node is relevant to its degree. Let the density of infected individuals at $n$ time step over the whole network be $\rho(n)$ numerically between $0$ and $1$. Intuitively, we have
\begin{equation}\label{eq:expk}
E[d_{j}(n)]=fk_{j}\rho(n).
\end{equation}
where $E[d_{j}(n)]$ is the expectation value of $d_{j}(n)$.

According to Eq. \ref{eq:expk}, if a node has a larger degree value, the more infected neighbors it may connect to, which rises higher risks of being infected.
As an instruction, Fig. \ref{fig:SIRS} explicitly presents the state transition of an individual in an epidemic based on the SIRS model.

\subsection{\label{sec:II2}Modeling on Population Size of Different Epidemic State}
As we know, the individual number in an epidemic state of a population varies with the evolving epidemic. Neglecting the birth and death of individuals in a short time scale, the fluctuation of the individual number in each epidemic state depend on the transitions between states and the migration of individuals. Besides, we also take the migration of mobile people into consideration. Queueing theory is introduced to explain the mechanism of transitions between states and fluctuations of the individual number of an epidemic state in detail. Furthermore, the number of individuals in all the three states in our epidemic model is regarded as a Markov Chain whose state space is $\{0, 1, 2, \cdots\}$.
\subsubsection{\label{que}Markov Queueing Network of SIRS}
\begin{figure}
\includegraphics[scale=0.298]{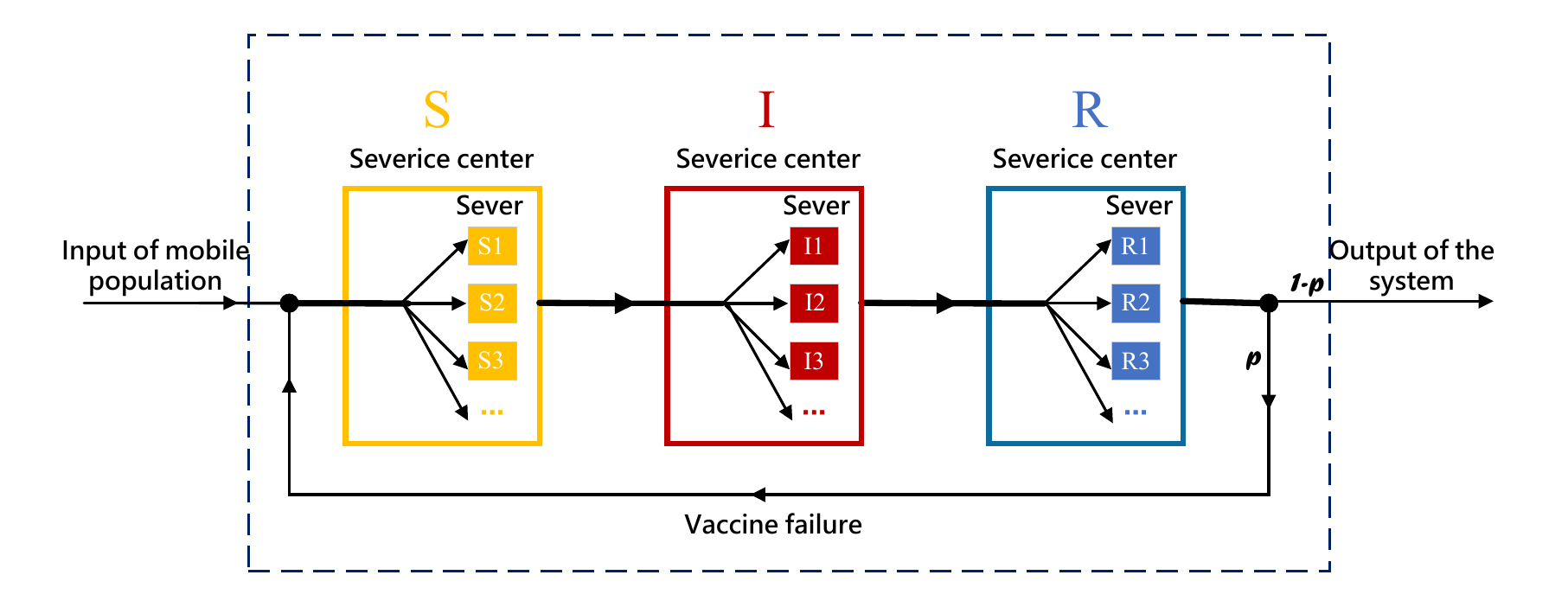}
\centering
\caption{An open Markov queueing network of SIRS model considering a mobile population: Inside the dotted box is the system(an area), where there are three solid line boxes representing service centers(an epidemic state is a service center). Each service center has infinite servers. Arrows indicate the transitions of individuals.}
\label{fig:queue}
\end{figure}

We hereby introduce the queueing theory to an SIRS epidemic model to describe the whole process explicitly. In a propagation process, there are transitions between epidemic states and also the migration among cities. Hence, we let a place, e.g., a city or a district be an open Markov network where three epidemic states are service centers. Each epidemic state is also described as a sub-system in which, if customers are under service, they remain staying in a state. Analyzing the input and output process of each state, the transition from $S$ to $I$ can be interpreted as the removal from $S$ while the arrival at $I$. Analogously, the transition from $I$ to $R$ is the removal from $I$ and as well as the arrival at $R$. The transition from $R$ to $S$ is part of the input of $S$ and the removal from $I$, since there are mobile individuals who arrive at $S$ outside the system and a certain portion of recovered individuals will leave the system rather than return to $S$. The infected rate is denoted by $\beta$, the recovered rate is $\gamma$, the input rate is $\lambda$, the output rate is $\alpha$, and the reviving proportion is $p$. For a better understanding, the queueing system is visually illustrated in Fig. \ref{fig:queue}.
We next construct the system in detail by defining its basic elements as follows.

\textbf{Input and output.}
The input people from outside the system can merely arrive at the susceptible service center and is regarded as a Poisson flow with the parameter $\lambda$. The output of the system stems from the decomposition of the Poisson output flow of the $R$ service center, which is a Poisson process as well.

\textbf{Mechanism of transitions.}
The input of the susceptible service center $S$ service center is a component of the input of mobile people from the outside and part of recovered individuals leaving $R$ and becoming susceptible again, which is a compound Poisson stream.
The input process of infected service center $I$ is a Poisson process with a variable rate since the input rate is related to the current number of customers under $I$ service center, which is called Queueing-Length-Dependent. The output is regarded as a Poisson stream as well.
The input flow of the infected service center $R$ is the output stream of the $I$ service center.

\textbf{Service mechanism.}
All service centers $S$, $I$ and $R$ are characterized by infinite servers numerically from $0$ to $\infty$. And the service time of each customer is independent identically distributed in all three systems. While the concrete time duration of service is different.
 The distribution service time of $S$ is taken as an exponential distribution with service rates related to the number of individuals in $I$ service center. The service time at $I$ service center is independent identically exponential distributed with the same parameter $\gamma$ of Poisson process.
The service time at $R$ service center is also independent identically exponential distributed with the same parameter $\alpha$ of Poisson process.

\textbf{Queueing Discipline.}
For our proposed queueing systems, by virtue of infinite servers,  new customers will accept service once it arrives at any service center. Consequently, the waiting time is avoided and the queueing discipline is trivial.

In the above mechanism, mobile people migrating to a place are all susceptible which is reasonable during the epidemic-prevent period, and the input number of people each time unit is random therefore following a Poisson distribution. While for the whole system, the output of it is all from the $R$ service center since only recovered people are allowed to leave the place. When an individual enters the system from the outside, it directly arrives at $S$ service center, then there is a probability that remains staying in $S$ or transferring to $I$ service center (be infected). An individual at $I$ has the probability to transfer to the $R$ service center. While individuals at $R$ have two choices, leaving the system or return to $S$ again.
The recovery of infected nodes occurs stochastically which suits the Poisson process, therefore, the input of $R$ system is a Poisson flow. The transition from $R$ to $S$ occurs randomly and the probability of occurrence is relatively small, the Poisson process with a small rate is therefore appropriate for that flow.
In terms of service mechanism, the number of individuals in an epidemic state is noticeable without limit and all coming individuals are allowed to join the service center. Hence the number of servers in each service center is endless.
Susceptible individuals being infected is related to the density of infected individuals. Accordingly, the staying time in $S$ service center depends on the number of customers under $I$ service center. Individuals keep in the $I$ and $R$ service center for some time and then leave, which occurs randomly, determining the service time at both the $I$ and $R$ follows an exponential distribution.

Then we discuss the underlying network where nodes represent individuals and edges between nodes indicate contact between individuals. We display the process of individuals joining in and leaving the system in the perspective of networks as follows.

\textbf{Initialization of networks.} The initial network is given an already constructed WS small-world network with a certain quantity of nodes defined by the scale $n_{0}$, the number of initial neighbors $k$, and the reconnection probability $p$.

\textbf{Connection and disconnection.} Mobile people floating to or leaving a city corresponds to the arriving and leaving process of nodes in the network. For any node coming with $m$ edges, it joins the network by $m$ edges connecting existing nodes. Assume the connection to a node occurs randomly and the nodes in the network are distributed randomly. The connection probability is equally $1/n$ for any node in the network where $n$ is the total number of nodes at that moment.

\textbf{Termination.} The terminal time is set to $T$ large enough for being stationary. The network ends evolving once the time reaches $T$.

The initial population network of an area can be regarded as a small-world network that is static and becomes dynamic because of migration. The above mechanism of a network describes that in the real situation, susceptible individuals enter an area and contact several existing individuals randomly, and recovered individuals can leave the area and break all their contacts. During this process, both of the structure and the scale of the network change with time.

Based on these indications, we next analyze the model focused on the number of individuals in three states.
\subsubsection{\label{mar}Numerical Study on Individual Number in Stationary}

As described above, the number of customers at three service centers can be respectively regarded as a continuous-time Markov Chain, $\{S(t), t\geq0\}$, $\{I(t), t\geq0\}$, $\{R(t), t\geq0\}$.
For all three epidemic states, the future number of individuals, given the present number and all past number of infected individuals depends only on the present number and independent of the past, satisfying $P\{I(t+h)=i|I(t)=j, I(u)=i(u), u\leq0<t\}=P\{I(t+h)=i|I(t)=j\}$ for $I(t)$, which is also suitable for $S(t)$ and $R(t)$. Then we take $\{S(t), t\geq0\}$, $\{I(t), t\geq0\}$, $\{R(t), t\geq0\}$ as Markov chains all with state space $\{0, 1, 2, \cdots\}$.

The initial condition is $\{S(0)=0, I(0)=1, R(0)=0\}$. And the susceptible individual number increases by 1 due to the revivification of a recovered individual becoming susceptible. A susceptible individual being infected leads to the decrease of susceptible number. Denote $P\{I(t+h)=i|I(t)=j\}$ by $P_{i,j}(t,h)$ as the probability that from state i at time $t$ to j at time $t+h$. Besides, since we study the epidemic model on homogeneous networks, we take the average degree denoted as $k$ of the network as the degree value of nodes. Then we have transition formulae demonstrated in Theorem\ref{l1}.

\begin{Theorem}\label{l1}
The transition probability of the susceptible individual number $S(t)$, the infected individual number $I(t)$ and the recovered individual number $R(t)$ are respectively
\begin{equation}\label{eq:s}
p^{S}_{ij}(t,h)=
\left\{
\begin{aligned}
&[\lambda+r(t)p{\alpha}]h+o(h),&j=i+1\\
&s(t)i(t)<k>{\beta}h+o(h),&j=i-1\\
&o(h), &\|j-i\|{\geq}2
\end{aligned}
\right.
,
\end{equation}
\begin{equation}\label{eq:i}
p^{I}_{ij}(t,h)=
\left\{
\begin{aligned}
&s(t)i(t)<k>{\beta}h+o(h), & j=i+1\\
&i(t){\gamma}h+o(h), &j=i-1\\
&o(t), &\|j-i\|{\geq}2
\end{aligned}
,
\right.
\end{equation}
and
\begin{equation}\label{eq:r}
p^{R}_{ij}(t,h)=
\left\{
\begin{aligned}
& i(t)p{\gamma}h+o(h),&j=i+1 \\
& r(t){\alpha}h+o(h),&j=i-1 \\
& o(h) &\|j-i\|{\geq}2
\end{aligned}
\right.
.
\end{equation}
where $s(t)$, $i(t)$, $r(t)$ respectively represents the number of susceptible, infected and recovered nodes at time $t$, $p$ is the ratio of recovered nodes entering $S$ state at time $t$.
\end{Theorem}
\begin{Proof}\normalfont
The input of $S$ service center is the Poisson flow of the migration of mobile individuals. It is along with part of the output of $R$ which also follows a Poisson process. The input of $S$ is a compound Poisson flow. Then, we first deduce that the probability of one node leaving $R$ and entering $S$ is $\sum_{m=1}^{r(t)}(e^{-{p\alpha}h})^{r(t)-1}{\cdot}(1-e^{-{p\alpha}h})=r(t)p{\alpha}h+o(h)$. The probability that one node come from outside the system is ${\lambda}he^{-{\lambda}h}$. One node coming to $S$ originate either outside the system or $R$. Therefore, in the light of Taylor's formula, the probability that one node enters $S$ is
\begin{equation}\label{eq:ins}
\begin{aligned}
P^{S}_{in}&=(e^{-{\lambda}t}+o(h)){\cdot}(r(t)p{\alpha}h+o(h))+{\lambda}he^{-{\lambda}h}{\cdot}(e^{-{\alpha}h})^{r(t)}\\
&=[1-{\lambda}h+o(h)][r(t)p{\alpha}h+o(h)]+{\lambda}h[1-{\lambda}h+o(h)]\\
&\cdot[1-p{\alpha}h]\\
&=[\lambda+pr(t)\alpha]h.
\end{aligned}
\end{equation}
Consequently, the input of $S$ can be regarded as a compound Poisson flow with the rate $\lambda+pr(t)\alpha$.
Then the probability that the service is unfinished is $e^{-i_{k}(t)k{\beta}h}$. In addition, the probability that $n$ nodes come to $S$ and exactly $n-1$ nodes leave is $o(h)$. Hence, according to above calculation and Eq.\ref{eq:ins}, we yield the transition probability of $S(t)$
\begin{equation}
\begin{aligned}
p^{S}_{i,i+1}&=(\lambda+r(t)\alpha)he^{-(\lambda+r(t)\alpha)h}{\cdot}e^{-i_{k}(t)k{\beta}h}+o(h)\\
&=(\lambda+r(t)\alpha)h[1-(\lambda+r(t)\alpha+i_{k}(t)k{\beta})h+o(h)]\\
&=(\lambda+r(t)\alpha)h+o(h).
\end{aligned}
\end{equation}
Analogously, the probability that none of the nodes become susceptible during $h$ is $e^{-(\lambda+r(t)\alpha)h}$. The probability of any node finishes the service and leaves $S$ is
\begin{equation}\label{eq:ous}
P^{S}_{out}=\sum_{m=1}^{s(t)}(e^{-i_{k}(t)k{\beta}h})^{s(t)-1}\cdot(1-e^i_{k}(t)k{\beta}h)=1-e^{i_{k}(t)k{\beta}h}.
\end{equation}
The probability that $n (n\geq2)$ nodes leave $S$ and exactly $n-1$ nodes come is $o(h)$. Together with Eq.\ref{eq:ous}, we obtain
\begin{equation}
\begin{aligned}
p^{S}_{i,i-1}&=e^{-(\lambda+r(t)\alpha)h}(1-e^{i_{k}(t)k{\beta}h})+o(h)\\
&=s_k(t)i_{k}(t)k{\beta}h+o(h).
\end{aligned}
\end{equation}
With the sum of probability being $1$,
\begin{equation}
p^{S}_{i,i}=1-(\lambda+r(t)\alpha+s_k(t)i_{k}(t)k{\beta})h+o(h).
\end{equation}

We next deduce the transition formula for $I(t)$.
The input of $I$ is the output of $S$ leading to the probability that one node comes to $I$ is $s(t)i(t)k{\beta}h+o(h)$, which is deduced above and the probability of none of nodes finish the service during time $h$ is $e^{-{\gamma}h}=1-{\gamma}h+o(h)$ yielding
\begin{equation}
\begin{aligned}
p^{I}_{i,i+1}&=[s(t)i(t)k{\beta}h+o(h)][1-{\gamma}h+o(h)]\\
&=s(t)i(t)k{\beta}h.
\end{aligned}
\end{equation}
The probability that one node finishes the service and leaves $I$ is $\sum^{i(t)}_{m=1}(e^{{\gamma}h})^{i(t)-1}{\cdot}(1-e^{{\gamma}h})=i(t){\gamma}h+o(h)$ and the probability that none of the nodes enter $I$ is equal to the probability that none of the nodes in the $S$ service center finish the service, which is noted as $e^{-i_{k}(t)k{\beta}h}$. Hence, we have
\begin{equation}
\begin{aligned}
p^{I}_{i,i-1}&=[i(t){\gamma}h+o(h)]e^{-i_{k}(t)k{\beta}h}\\
&=[i(t){\gamma}h+o(h)][1-i_{k}(t)k{\beta}h+o(h)]\\
&=i(t){\gamma}h+o(h).
\end{aligned}
\end{equation}
The rest demonstration of the transition formula of $R(t)$ is essentially the same as the above proof.

Above all, the results follow.
\end{Proof}
Th.\ref{l1} presents the transition probability of the individual number of three states. It is worth noting that the transition probability is not the epidemic transition probability but the probability of the population size changing.

\begin{Theorem}\label{ty2}
The expectation of the population in three states under stationary is respectively
$E[S]=\frac{\gamma}{\beta k}$, $E[I]=\frac{\lambda}{\gamma(1-p)}$, $E[R]=\frac{\lambda}{\alpha(1-p)}$.
\end{Theorem}
\begin{Proof}\normalfont
When the process goes in a stationary regime, the input rate and the output rate of a state are equal. Suppose the expectation value of the number of susceptible, infected and recovered individuals under stationary are $E[S]$, $E[I]$ and $E[R]$. According to Th. \ref{l1}, we get the following equilibrium equations:
\begin{equation}\label{eq:ex}
\left\{
\begin{aligned}
\lambda+p\alpha E[R]&=E[S]E[I]\beta k\\
E[S]E[I]\beta k&=\gamma E[I]\\
\gamma E[I]&=\alpha E[R]
\end{aligned}
\right.
.
\end{equation}
Solving Eq. \ref{eq:ex}, we attain the expectation of the number of individuals of the three states.

The results follow.
\end{Proof}

Hence, we come to the following important conclusion.
Eq. \ref{eq:ex} combined with Th. \ref{l1} reveals the relation between the change of the population size in three states and the transition among the states. In detail, the number of individuals in the susceptible state decreases by 1 indicating one individual transfers from the susceptible state to the infected state, while it increases by one because of an arrival which is the compound Poisson flow composed of the migration to the area and part of recovered individuals turning susceptible. For the infected state, the population decreases when an infected individual transforms into a recovered state. The infected population increases by one owing to a transition from the susceptible to the infected state. Analogously, a leave from the recovered state contributes to one decrease of population, and a transition from infected state leads to an increase.

According to the above theoretical results, we will carry out real experiments to further verify our model.
\section{\label{sec:III}Simulation}
\begin{figure}[!t]
\includegraphics[width=8.3cm, height=5.8cm]{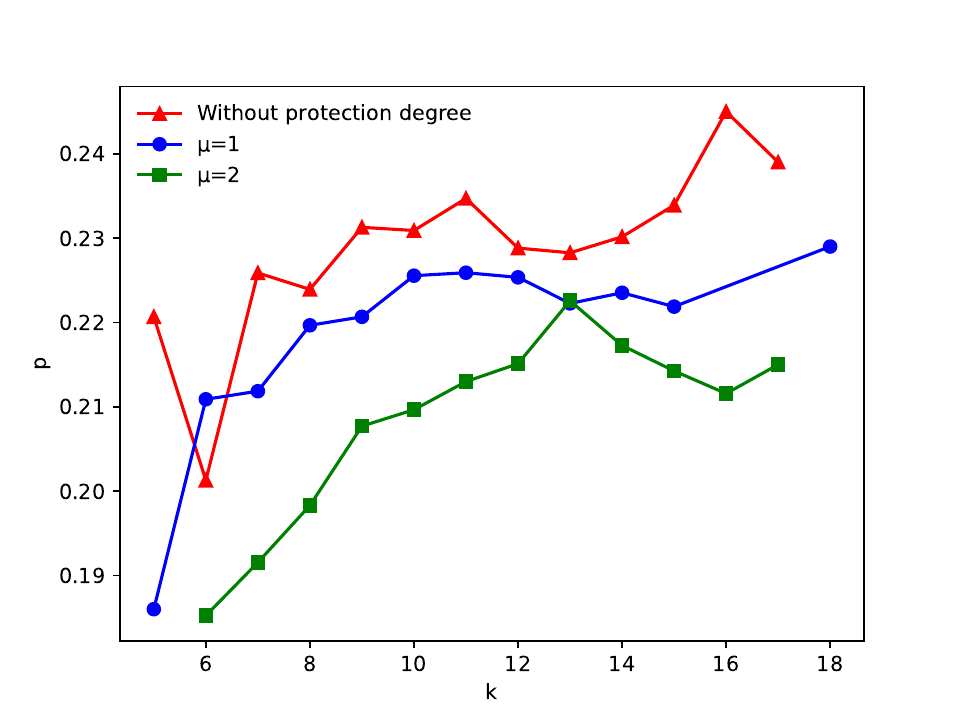}
\centering
\caption{The infected probabilities with different protection degrees: We take the average value of the infected probability of individuals within the same-degree group. The protection degree with $\mu=1$, $\mu$=2, and without protection degree is respectively denoted by the blue circle, the green square and the red triangle plot. The protection degree lower the probability of being infected, and larger value of $\mu$ leads to smaller infected probability.}
\label{fig:prok}
\end{figure}
In this section, we simulate the propagation process based on proposed models. In the first sub-section, we simulate the transition among epidemic states according to the individual model and investigate the relation between the limited distribution of epidemic states and the degree of nodes. The second sub-section based on the whole population model showcases the properties of the system, e.g., the number of individuals and the limited distribution of the number of individuals in $S, I, R$ respectively.
\subsection{Transition of Epidemic State}\label{subsection:idvd}
The limited distribution of state $S, I, R$ indicates the probability of individuals being at $S, I, R$ correspondingly. In other words, it can be regarded as the proportion of the number of individuals in three epidemic states. We reveal that the propagation process will be stationary and the limited probability of being infected is related to the degree of a node standing and also the protection degree of an individual.
Moreover, according to the transition matrix, the propagation process is decided by the infected rate, recovered rate, the revived rate, and especially the protection degree, hence we set different values for the parameter of the Poisson distribution $\mu$.

The underlying network is a WS small-world network whose distribution of degree follows a Poisson distribution and the average path length is short, thus it fits the population network in an area. A WS small-world network is constructed by randomly reconnecting a regular network in which each node has $k$ neighbors both on the left and right sides. In our experiments, the network is initialized by $networkx$ in Python, where the scale of the network is set to be 1000, each vertex is connected to 5 neighbors on the left and right respectively, and the reconnection probability is 0.5. And we utilize $numpy.random.poisson()$ function to generate protection degrees. The experiments are carried out by setting different values 1 and 2 for the parameter of Poisson distribution and we make a comparison with the propagation process where individuals without protection degrees. The terminal time is set to be 1500 that is long enough to let the whole process be stationary. As time passes by, we record each individual's state at each time step when the system is stationary. Eventually, we calculate the mean value of frequencies of being at I states of nodes classified by degree $k$ to evaluate the level of epidemics.
\begin{figure}[!t]
\includegraphics[width=8.3cm, height=5.8cm]{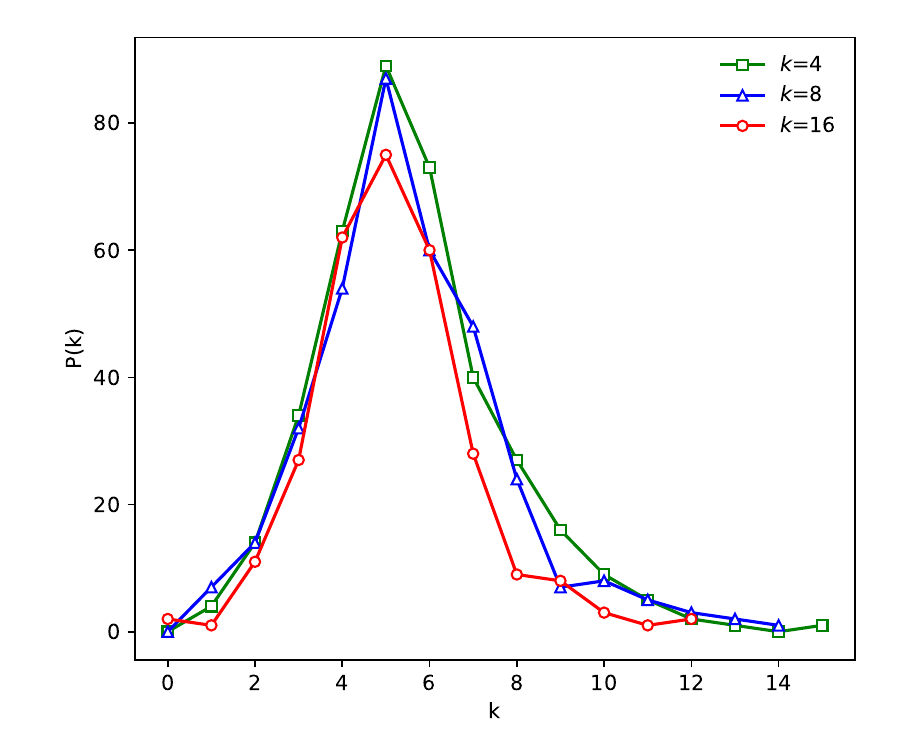}
\centering
\caption{The degree distribution of evolved network with different initial edges of each node: The edges of each node in the initial regular network are set to be 4, 8, 16 and the node joining the network with edges $m=4$. The stationary distributions are identical to each other though the initial edges is different.}
\label{fig:enwkdd}
\end{figure}

Since networks are generated randomly in each experiment, the degree are not the same as we can see from Fig. \ref{fig:prok}, the lines have different degree values on the horizontal coordinate. Hence, we focus on the mean values of infected rate of degree between 6 and 16. The red triangle plot indicates the limited probability of being infected without the protection degree which lays above the other two lines, which indicate the individual protection degree deduce the probability of being infected. The blue circle plot is the infected probability of individuals with the protection degree following a Poisson distribution whose parameter is 1 while the green square plot is the infected probability of individuals with Poisson parameter 2. The green plot lays below the green one, indicating that larger intensity of Poisson distribution for protection degree reduces the limited infected probability. Furthermore, three plots have an upward tendency as the degree value increases, which indicates a larger degree value causes a higher probability of being infected. Consequently, the simulation results can give suggestions that a higher level of protective awareness among people leads to a lower risk of being infected and fewer contacts to people also decrease the infected probability.
\subsection{Queueing System of SIRS}
\begin{figure*}[!t]
\begin{minipage}[t]{0.33\linewidth}
\centering
 \includegraphics[height=3.55cm, width=5.7cm]{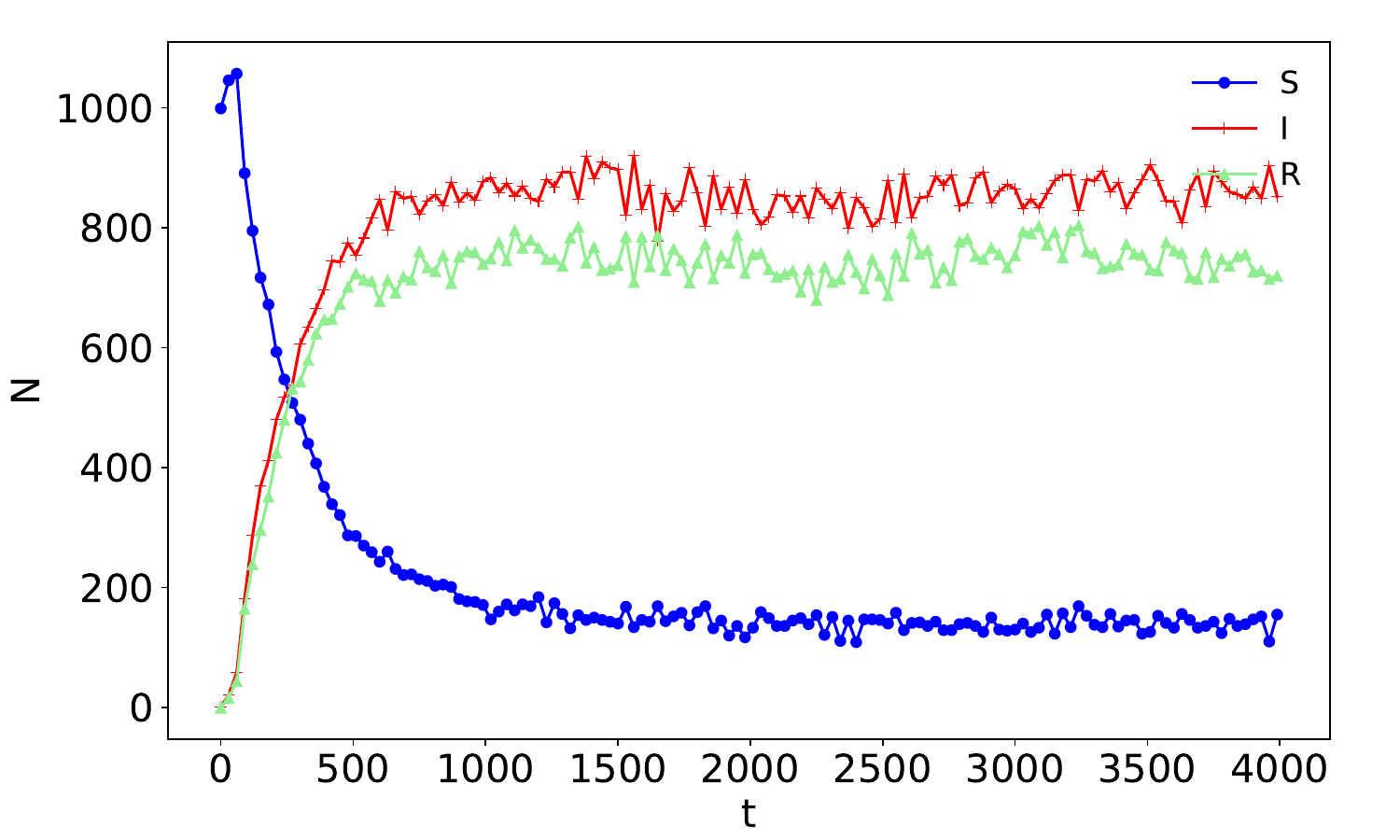}
\parbox{4.5cm}{\centering\footnotesize \hspace{0.1cm}(a) The individual number with the initial setting of parameters}

\end{minipage}
\begin{minipage}[t]{0.33\linewidth}
\centering
 \includegraphics[height=3.55cm, width=5.7cm]{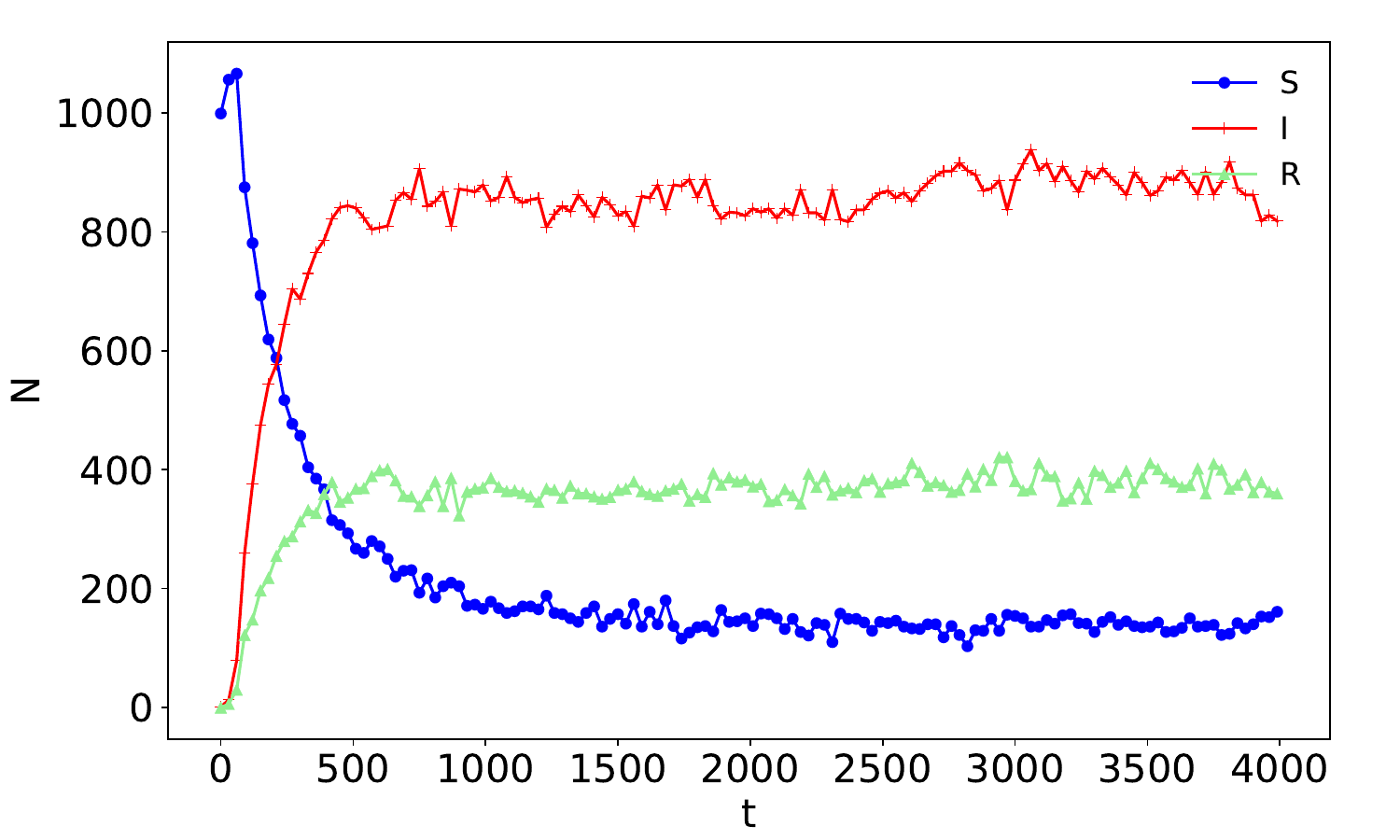}
\parbox{4.5cm}{\centering\footnotesize \hspace{0.1cm}(b) The individual number with $\alpha$ adjusted to 1.6}

\end{minipage}
\begin{minipage}[t]{0.33\linewidth}
\centering
 \includegraphics[height=3.55cm, width=5.7cm]{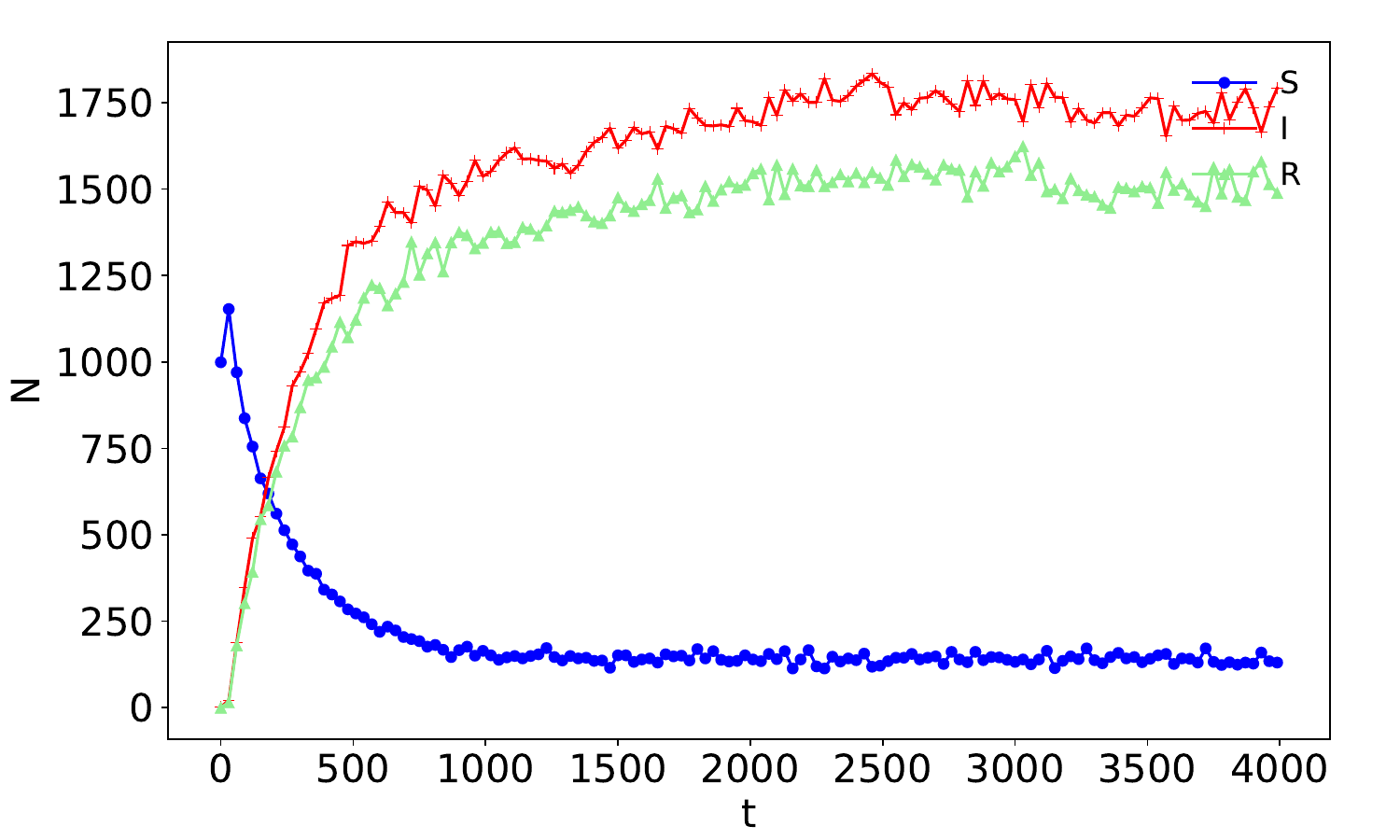}
\parbox{4.5cm}{\centering\footnotesize \hspace{0.1cm}(c) The individual number with $\lambda$ adjusted to 6}

\end{minipage}
\begin{minipage}[t]{0.33\linewidth}
\centering
 \includegraphics[height=3.55cm, width=5.7cm]{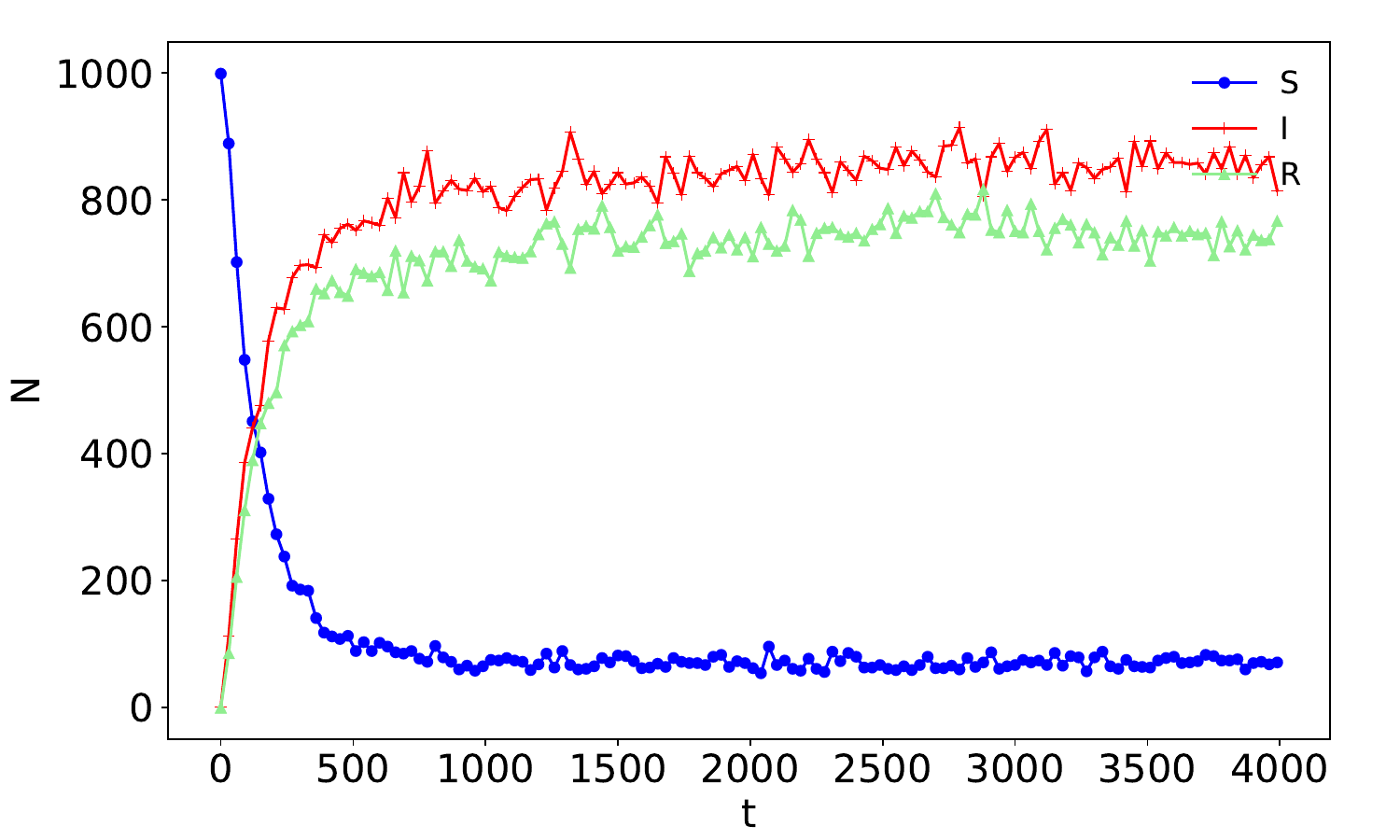}
\parbox{4.5cm}{\centering\footnotesize \hspace{0.1cm}(d) The individual number with $\beta$ adjusted to 0.002}

\end{minipage}
\begin{minipage}[t]{0.33\linewidth}
\centering
 \includegraphics[height=3.55cm, width=5.7cm]{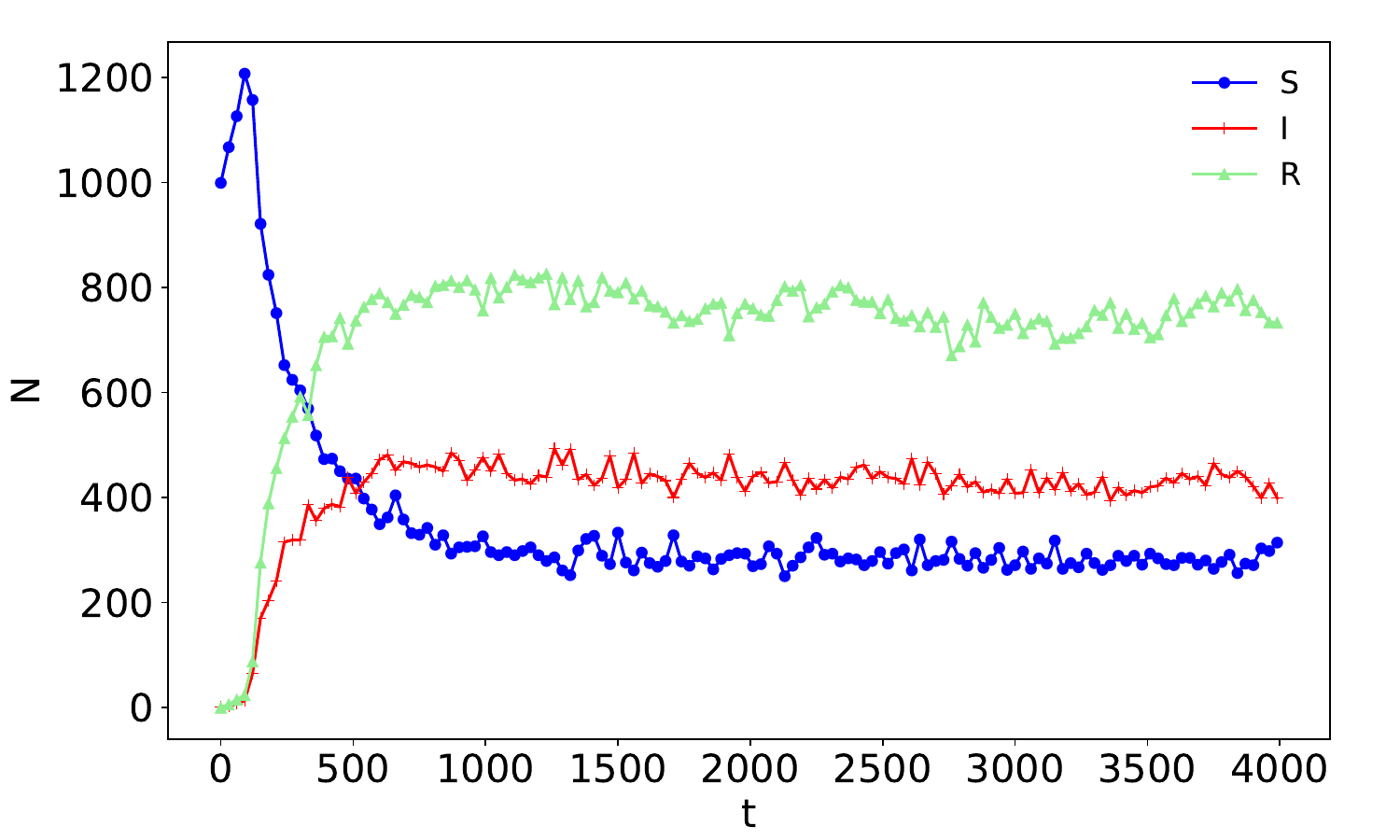}
\parbox{4.5cm}{\centering\footnotesize \hspace{0.1cm}(e) The individual number with $\gamma$ adjusted to 1.4}

\end{minipage}
\begin{minipage}[t]{0.33\linewidth}
\centering
 \includegraphics[height=3.55cm, width=5.7cm]{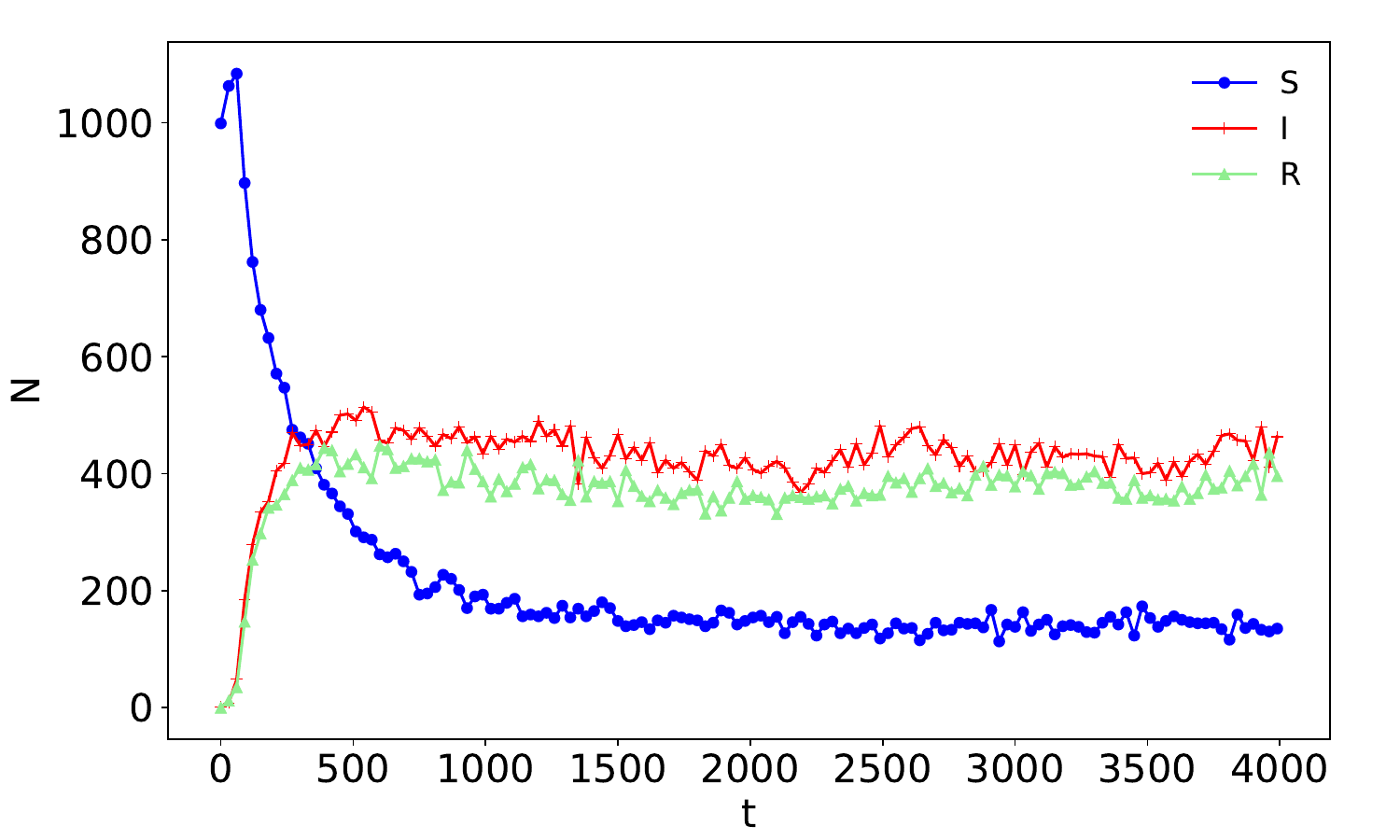}
\parbox{4.5cm}{\centering\footnotesize \hspace{0.1cm}(f) The individual number with $p$ adjusted to 0.990}

\end{minipage}
\caption{\label{fig:n123}The number of individuals varying with time of the three states with different parameters: The figure respectively shows the influence of each parameter on the individual number. Sub-Fig. (a) is set as $\lambda$=3, $\beta$=0.001, $\gamma$=0.7, $\alpha$=0.8 and $p$=0.995. $\alpha$, $\lambda$, $\beta$, $\gamma$ and $p$ are changed respectively from (b)-(e).}
\end{figure*}

\begin{table*}[!t]
  \centering
  \caption{The results of the stationary individual number of $S$ with $\lambda$=3, $\alpha$=0.8, $p$=0.995, $<k>$=5.}\label{tab:S}
    \begin{tabular}{lccc}
    \hline
    \hline
    \multirow{2}{*} & \multicolumn{3}{c}{Parameters ($\beta$, $\gamma$)}\\
    \cline{2-4}
       Results   & (0.001, 0.7) & (0.002, 0.7) & (0.001, 1.4) \\
    \hline
    Theoretical results & 140   & 70    & 280 \\
    Simulation results & 141   & 69    & 285 \\
    Standard deviation & 0.123 & 0.119 & 0.119 \\
    Relative error & 0.7\% & 1.4\% & 1.8\% \\
    \hline
    \hline
    \end{tabular}%
  \label{tab:addlabel}%
\end{table*}%

\begin{table*}[!t]
  \centering
  \caption{The results of the stationary individual number of $I$ with $\beta$=0.001, $\alpha=0.8$ and $<k>$=5.}\label{tab:I}
    \begin{tabular}{lcccc}
    \hline
    \hline
    \multirow{2}{*} & \multicolumn{3}{c}{Parameters ($\lambda$, $\gamma$, $p$)}\\
    \cline{2-5}
       Results   & (3, 0.7, 0.995) & (6, 0.7, 0.995) &  (3, 1.4, 0.995) & (3, 0.7, 0.990) \\
    \hline
    Theoretical results & 857   & 1714  & 429   & 429 \\
    Simulation results & 841   & 1746  & 419   & 444 \\
    Standard deviation & 0.125 & 0.127 & 0.128 & 0.124 \\
    Relative error & 1.9\% & 1.9\% & 2.3\% & 3.5\% \\
\hline
\hline
    \end{tabular}%
  \label{tab:addlabel}%
\end{table*}%
\begin{table*}[!t]
  \centering
  \caption{The results of the stationary individual number of $R$ with $\beta$=0.001, $\gamma=0.7$ and $<k>$=5.}\label{tab:R}
    \begin{tabular}{lcccc}
    \hline
    \hline
    \multirow{2}{*} & \multicolumn{3}{c}{Parameters ($\lambda$, $\alpha$, $p$)}\\
    \cline{2-5}
       Results   & (3, 0.8, 0.995)& (6, 0.8, 0.995) & (3, 1.6, 0.995) & (3, 0.8, 0.990) \\
    \hline
    Theoretical results & 750   & 1500  & 375   & 375 \\
    Simulation results & 737   & 1529  & 379   & 389 \\
    Standard deviation & 0.125 & 0.123 & 0.119 & 0.131 \\
    Relative error & 1.7\% & 1.9\% & 1.1\% & 3.7\% \\
    \hline
    \hline
    \end{tabular}%
  \label{tab:addlabel}%
\end{table*}%

\begin{figure*}[htbp]
\begin{minipage}[!t]{0.33\linewidth}
\centering
 \includegraphics[height=4.4cm, width=6.6cm]{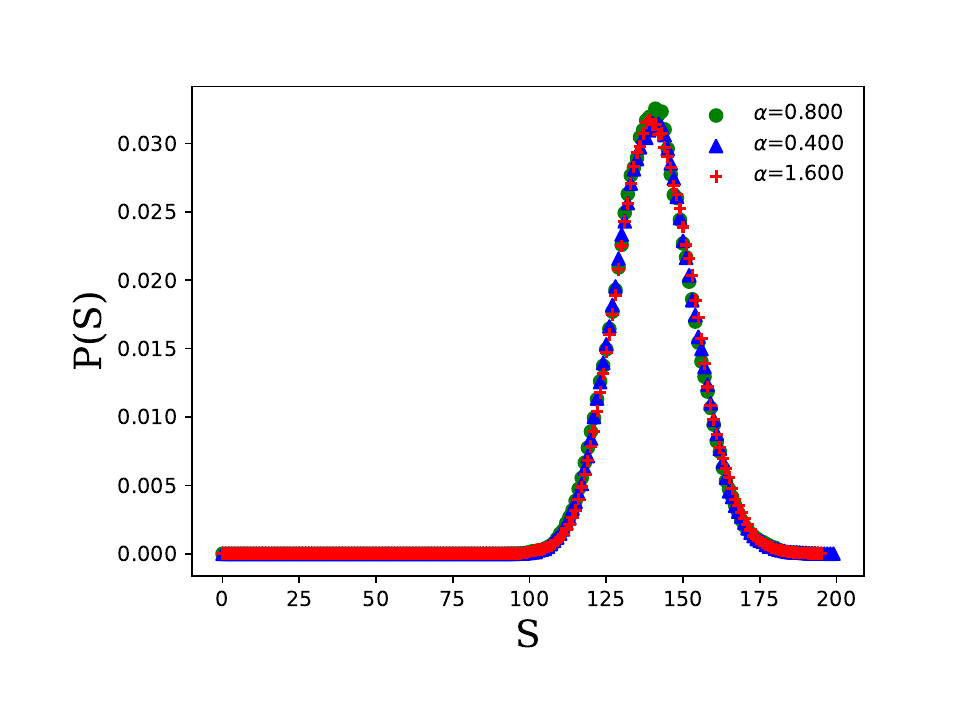}
\parbox{4.4cm}\centering{\footnotesize \hspace{0.1cm}(a) The distribution of $S$ individual number with different values of $\alpha$}
\end{minipage}
\begin{minipage}[!t]{0.33\linewidth}
\centering
 \includegraphics[height=4.4cm, width=6.6cm]{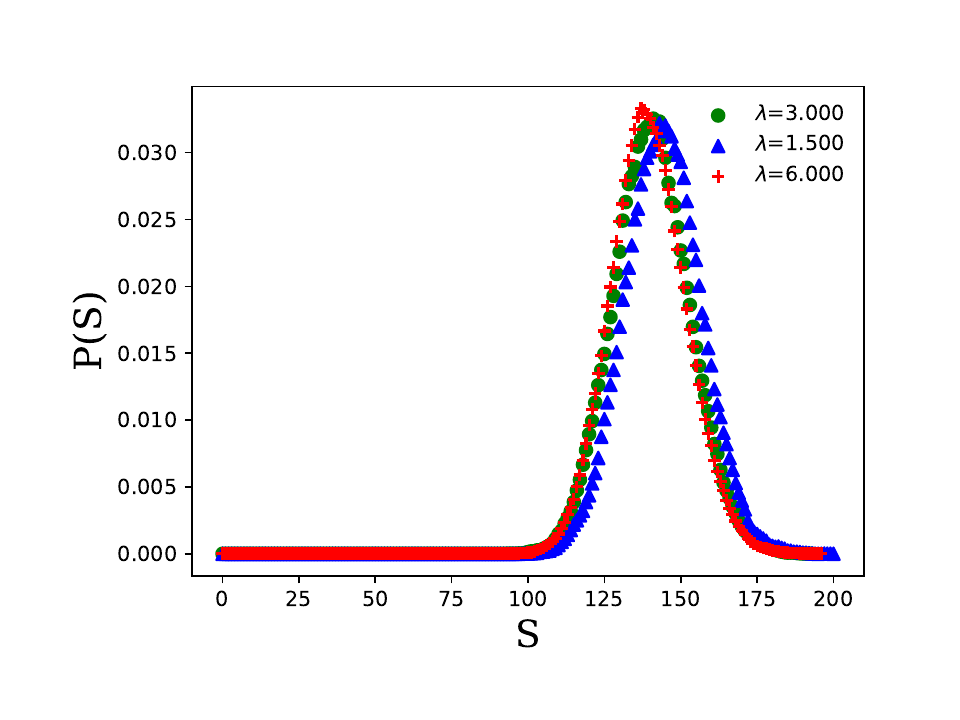}
\parbox{4.4cm}\centering{\footnotesize \hspace{0.1cm}(b) The distribution of $S$ individual number with different values of $\lambda$}
\end{minipage}
\begin{minipage}[!t]{0.33\linewidth}
\centering
 \includegraphics[height=4.4cm, width=6.6cm]{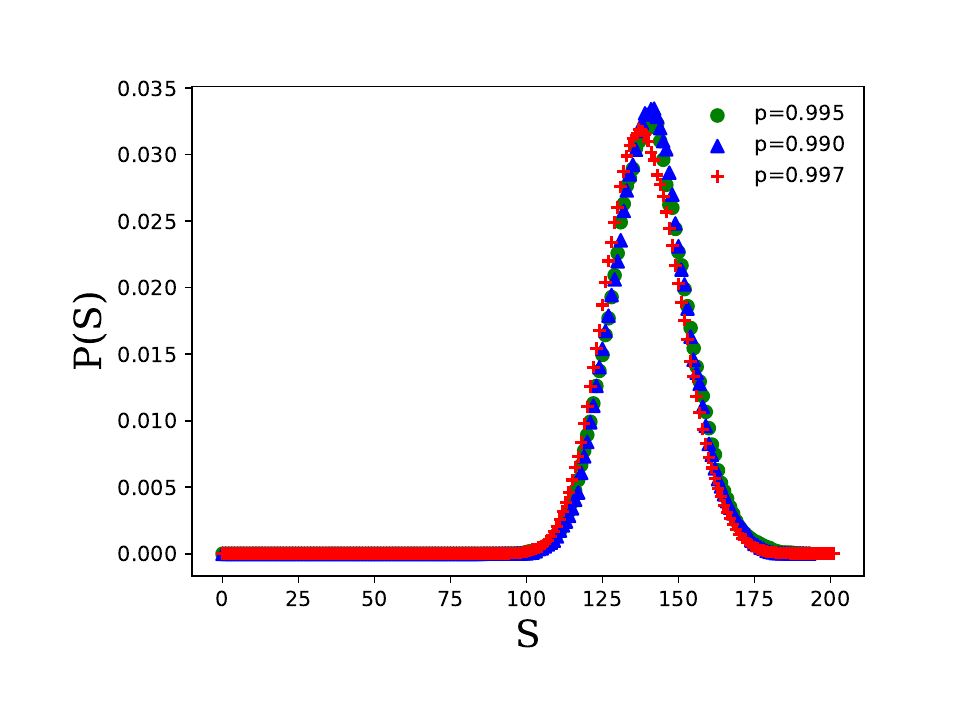}
\parbox{4.4cm}\centering{\footnotesize \hspace{0.1cm}(c) The distribution of $S$ individual number with different values of $p$}
\end{minipage}
\begin{center}
\begin{minipage}[!t]{0.34\linewidth}
\centering
 \includegraphics[height=4.4cm, width=6.6cm]{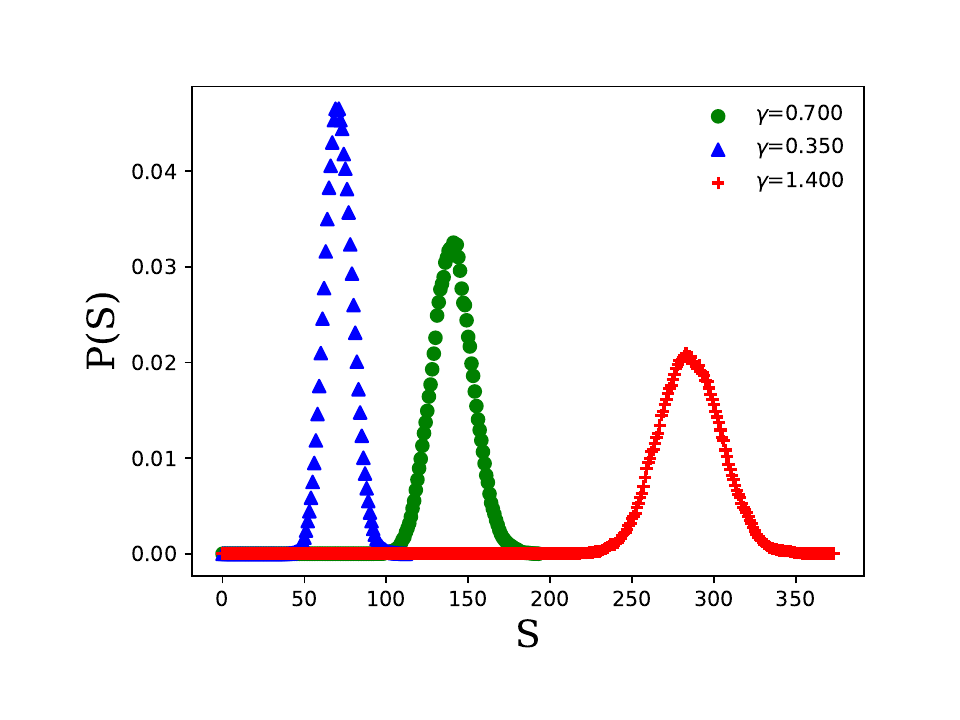}
\parbox{8.5cm}{\footnotesize\hspace{0.5cm}\vspace{0.15cm}{(d) The distribution of $S$ individual number with \par \qquad \qquad \qquad  \qquad different values of $\gamma$}}
\end{minipage}
\begin{minipage}[!t]{0.34\linewidth}
\centering
 \includegraphics[height=4.4cm, width=6.6cm]{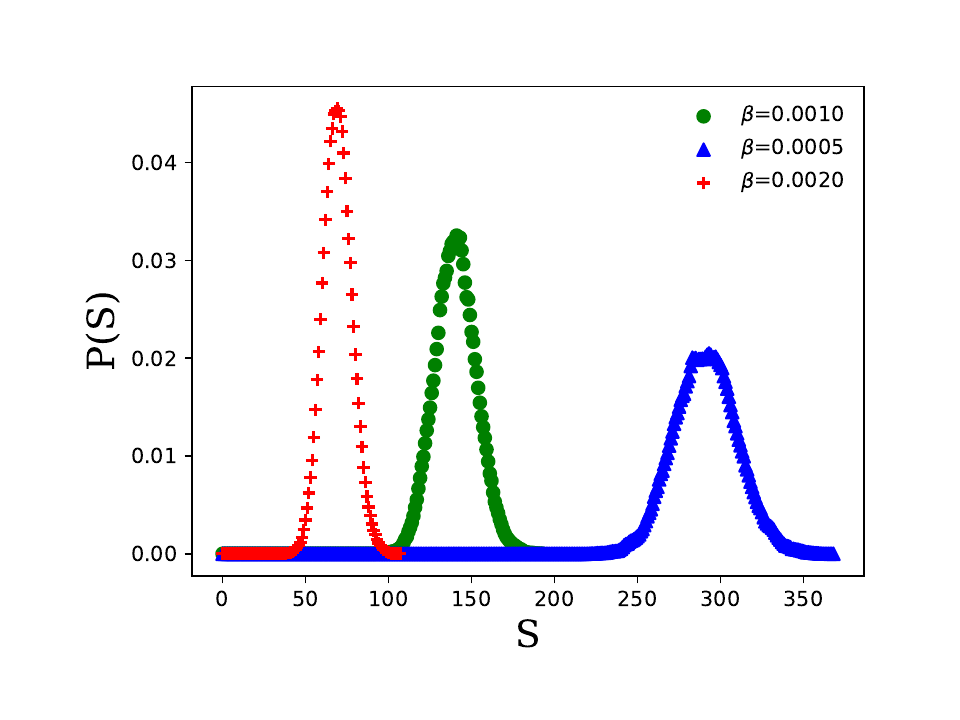}
\parbox{8.5cm}{\footnotesize \hspace{0.5cm}\vspace{0.15cm}{(e) The distribution of $S$ individual number with  \par \qquad \qquad \qquad  \qquad different values of $\beta$}}
\end{minipage}
\end{center}

\caption{\label{fig:ds}The comparison of the $S$ individual number distributions with different parameters: Five parameters influence the distribution of the individual number of $S$ differently. $\gamma$ and $\beta$ are deciding factors of $S$, where $\gamma$ is proportional to the individual number while $\beta$ works inversely. The initial setting is $\lambda$=3, $\beta$=0.001, $\gamma$=0.7, $\alpha$=0.8 and $p$=0.995. Sub-Figs. (a)-(e) display the distributions based on the original parameters changing $\alpha$, $\lambda$, $p$, $\gamma$ and $\beta$ orderly.}
\end{figure*}

\begin{figure*}[htbp]
\begin{minipage}[H]{0.33\linewidth}
\centering
 \includegraphics[height=4.4cm, width=6.6cm]{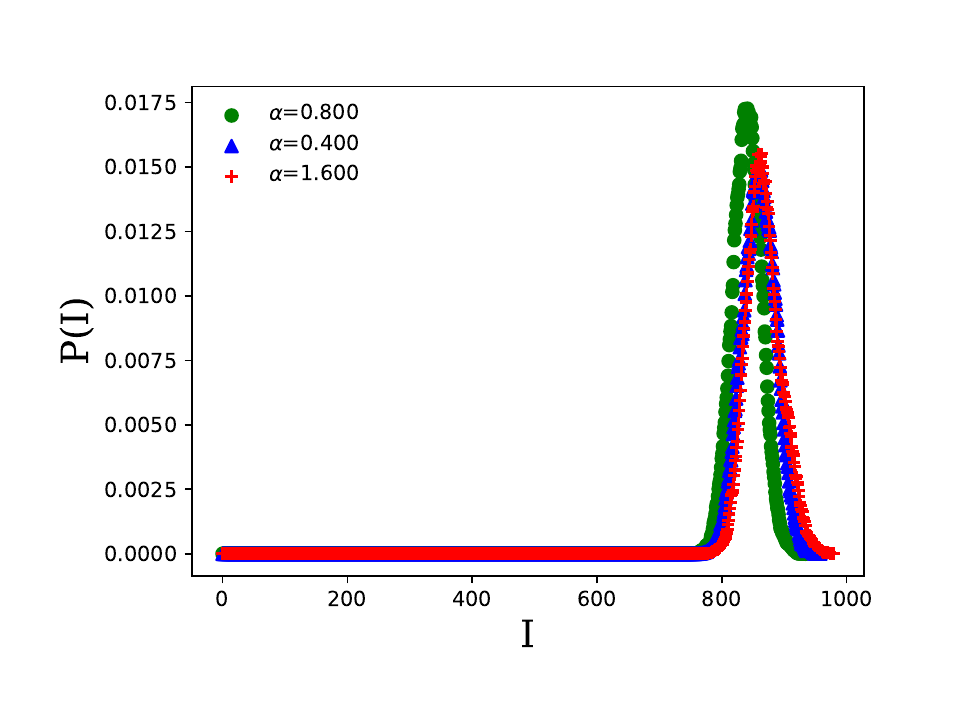}
\parbox{4.4cm}\centering{\footnotesize \hspace{0.1cm}(a) The distribution of $S$ individual number with different values of $\alpha$}
\end{minipage}
\begin{minipage}[H]{0.33\linewidth}
\centering
 \includegraphics[height=4.4cm, width=6.6cm]{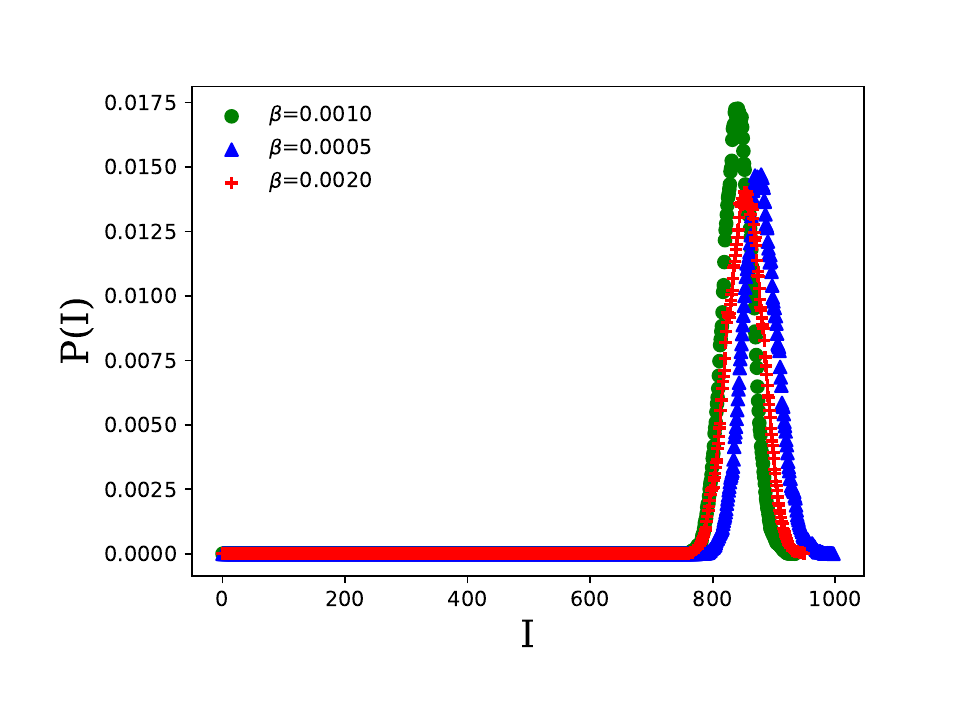}
\parbox{4.4cm}\centering{\footnotesize \hspace{0.1cm}(b) The distribution of $S$ individual number with different values of $\beta$}
\end{minipage}
\begin{minipage}[H]{0.33\linewidth}
\centering
 \includegraphics[height=4.4cm, width=6.6cm]{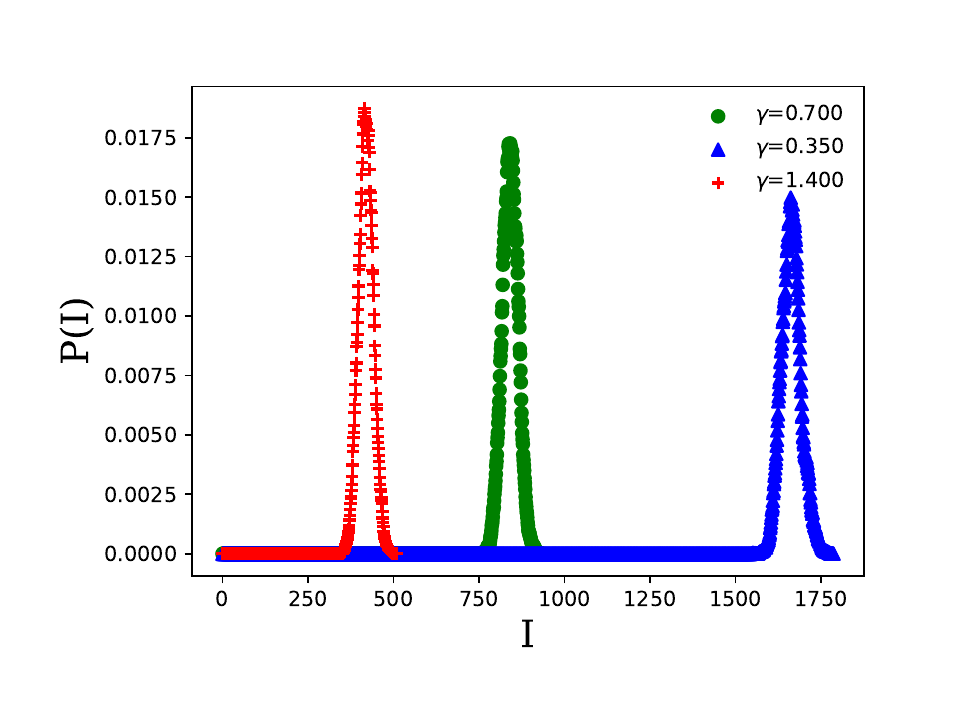}
\parbox{4.4cm}\centering{\footnotesize \hspace{0.1cm}(c) The distribution of $S$ individual number with different values of $\gamma$}
\end{minipage}
\begin{center}
\begin{minipage}[H]{0.34\linewidth}
\centering
 \includegraphics[height=4.4cm, width=6.6cm]{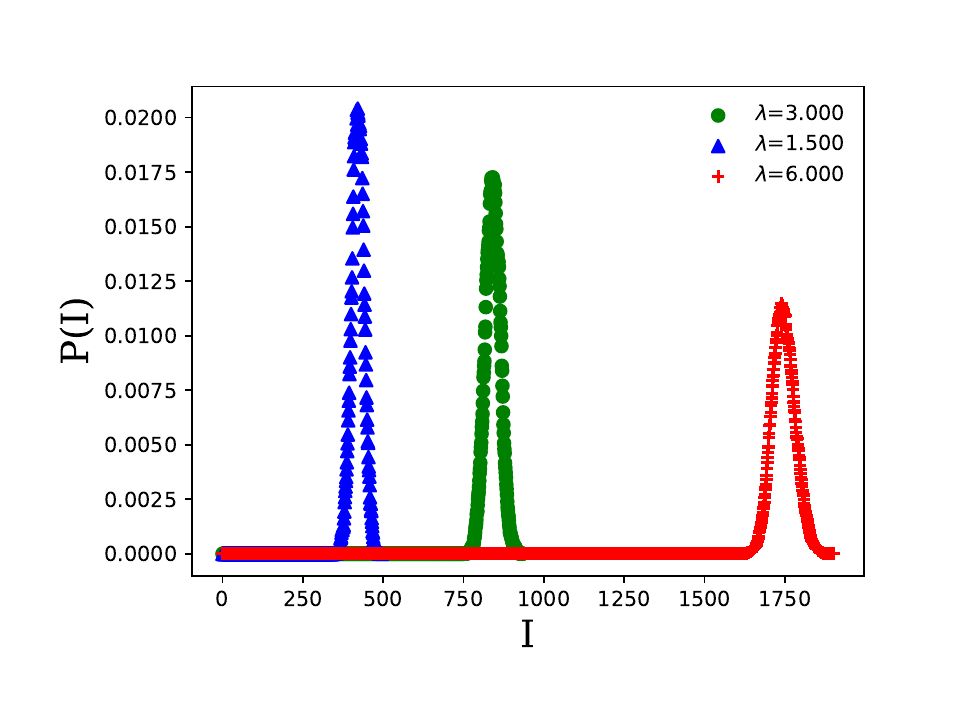}
\parbox{8.5cm}{\footnotesize\hspace{0.5cm}\vspace{0.15cm}{(d) The distribution of $S$ individual number with \par \qquad \qquad \qquad  \qquad different values of $\lambda$}}
\end{minipage}
\begin{minipage}[H]{0.34\linewidth}
\centering
 \includegraphics[height=4.4cm, width=6.6cm]{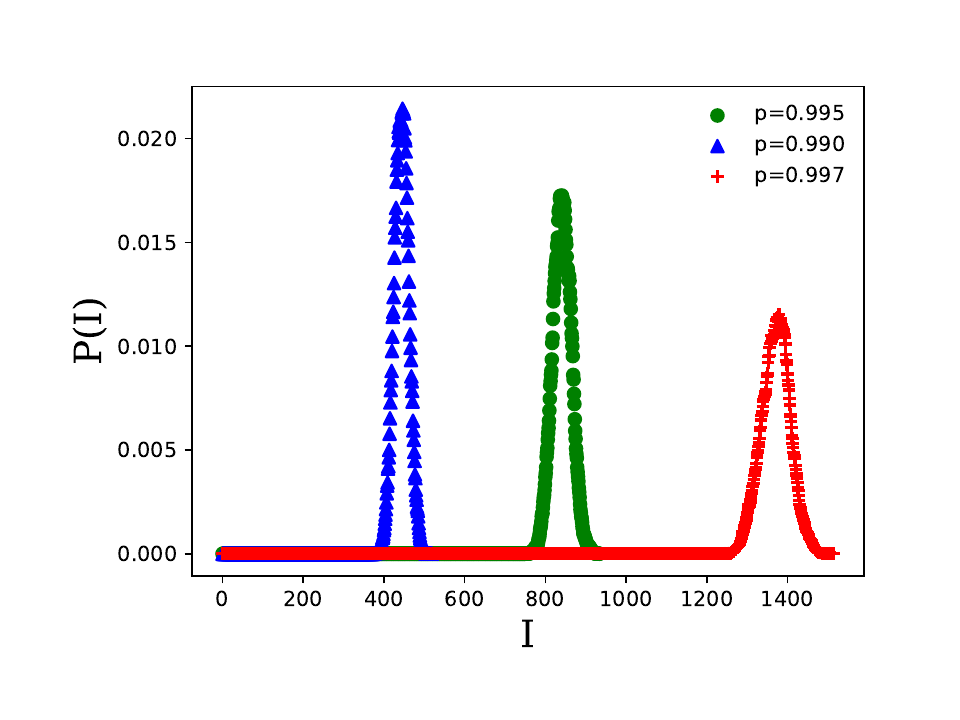}
\parbox{8.5cm}{\footnotesize \hspace{0.5cm}\vspace{0.15cm}{(e) The distribution of $S$ individual number with  \par \qquad \qquad \qquad  \qquad different values of $p$}}
\end{minipage}
\end{center}
\caption{\label{fig:di}The comparison of the $I$ individual number distribution with different parameters: Five parameters influence the distribution of the individual number of $S$ differently where $\lambda$ and $p$ are proportional to the individual number while $\gamma$ is inversely proportional to that. The initial setting is $\lambda$=3, $\beta$=0.001, $\gamma$=0.7, $\alpha$=0.8 and $p$=0.995. Sub-Figs.(a)-(e) display the distributions based on the original parameters changing $\alpha$, $\gamma$, $\lambda$, $\beta$ and $p$ orderly.}
\end{figure*}

The following simulations concentrate on the number of individuals in each epidemic state. We first present the varying number of three states with time going where the system will be stationary when time is large enough. Besides, we illustrate the individual number distribution with different parameters. In these experiments, we apply $np.random.exponential$ function to generate time series as the staying time for individuals in three epidemic states. And we update the number of individuals $S(t)$, $I(t)$, and $R(t)$ whenever an event occurs. Time is set to be large enough for stationarity.

Primarily, we demonstrate the degree distributions of the evolving network. The initial network is a small-world network described above in Sec. \ref{subsection:idvd}. In the following experiments, the initial scale of the network is set to be 1000, a number of nearest neighbors $k$ is set to be 4 and the reconnection probability $p$ is set to be 0.7. To analyze the influence of $k$ on the output network constructed by our proposed model whose edge number $m$ of a newly arriving node coming is set to be 4, we let the initial parameter $k$ be 4, 8, and 16 and present the degree distributions $P(k)$ in Fig. \ref{fig:enwkdd}. As we can see, the degree distributions of three networks with different initial edges have the same mode and they are all still homogenous. Furthermore, the average degree calculated is 5 via the degree distribution when the network is stationary, which demonstrates that the distribution of the output network is independent of the initial number of edges $k$.

\begin{figure*}[htbp]
\begin{minipage}[H]{0.33\linewidth}
\centering
 \includegraphics[height=4.4cm, width=6.6cm]{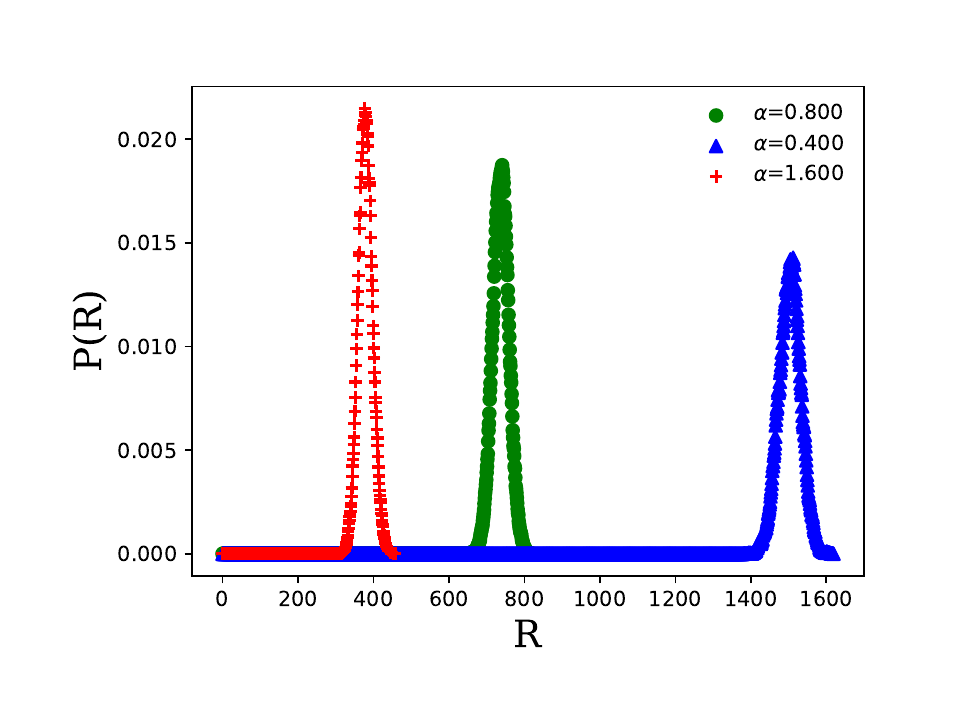}
\parbox{4.4cm}\centering{\footnotesize \hspace{0.1cm}(a) The distribution of $S$ individual number with different values of $\alpha$}

\end{minipage}
\begin{minipage}[H]{0.33\linewidth}
\centering
 \includegraphics[height=4.4cm, width=6.6cm]{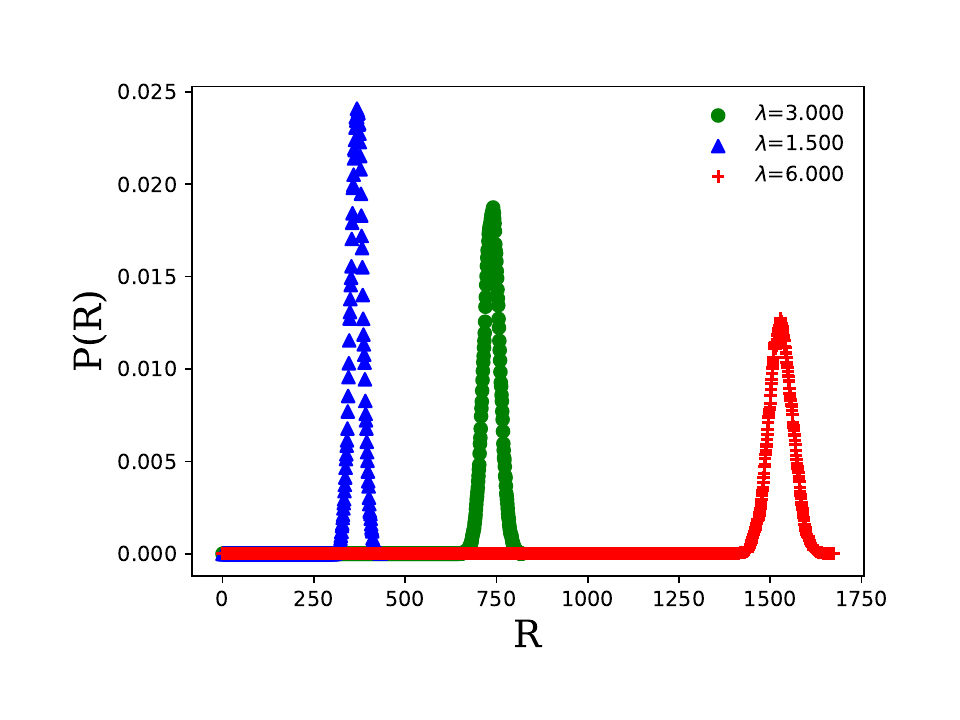}
\parbox{4.4cm}\centering{\footnotesize \hspace{0.1cm}(b) The distribution of $S$ individual number with different values of $\lambda$}

\end{minipage}
\begin{minipage}[H]{0.33\linewidth}
\centering
 \includegraphics[height=4.4cm, width=6.6cm]{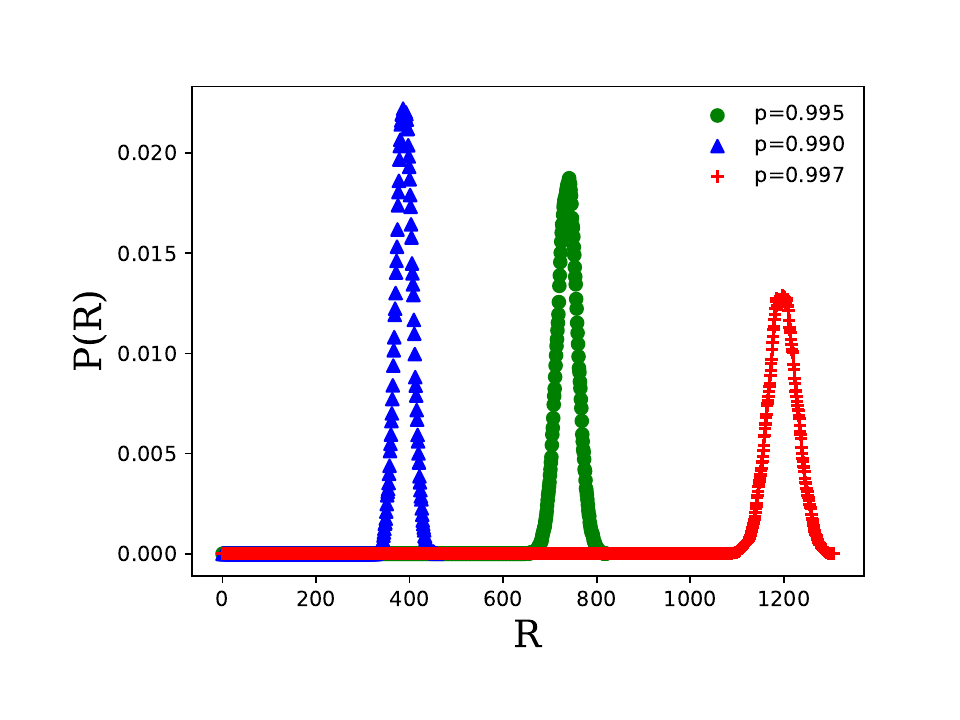}
\parbox{4.4cm}\centering{\footnotesize \hspace{0.1cm}(c) The distribution of $S$ individual number with different values of $p$}

\end{minipage}

\begin{center}
\begin{minipage}[H]{0.34\linewidth}
\centering
 \includegraphics[height=4.4cm, width=6.6cm]{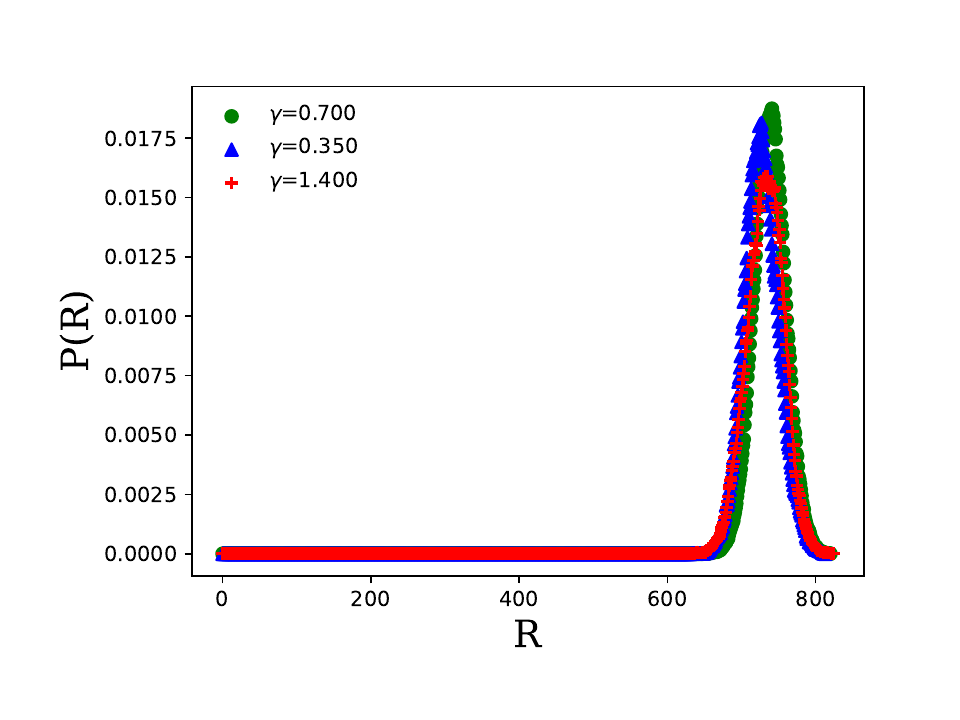}
\parbox{8.5cm}{\footnotesize \hspace{0.5cm}\vspace{0.15cm}(d) The distribution of $S$ individual number with \par \qquad \qquad \qquad  \qquad different values of $\gamma$}

\end{minipage}
\begin{minipage}[h]{0.34\linewidth}
\centering
 \includegraphics[height=4.4cm, width=6.6cm]{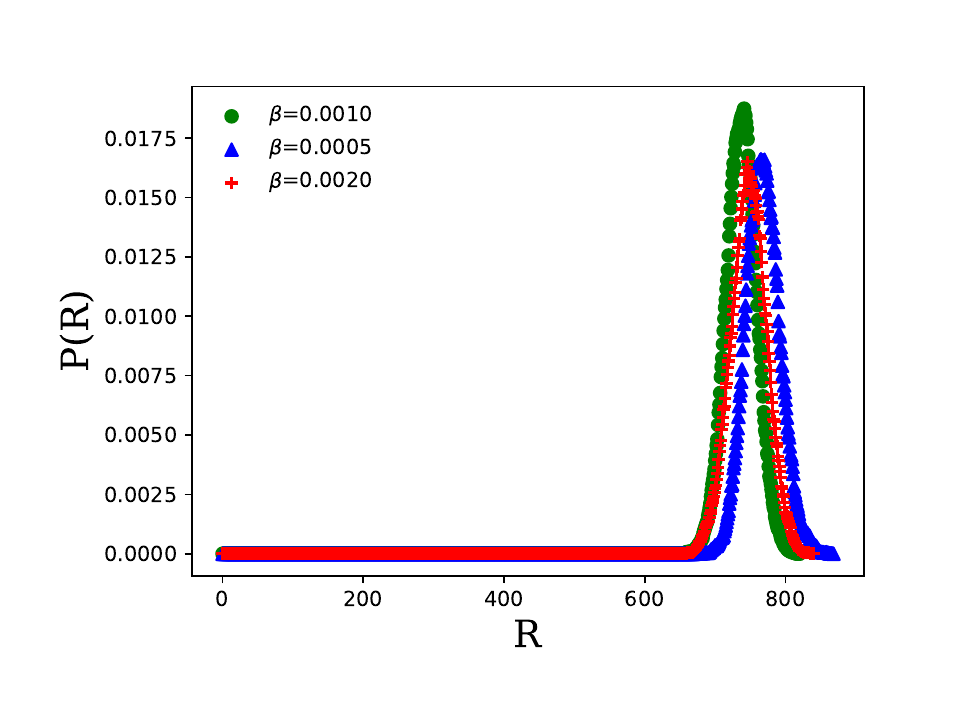}
\parbox{8.5cm}{\footnotesize \hspace{0.5cm}\vspace{0.15cm}(e) The distribution of $S$ individual number with \par \qquad \qquad \qquad  \qquad different values of $\beta$}

\end{minipage}
\end{center}
\caption{\label{fig:dr}The comparison of the $R$ individual number distribution with different parameters: Five parameters influence the distribution of the individual number of $S$ differently. $\lambda$ and $p$ are proportional to the individual number while $\alpha$ has an inverse effect. The initial setting is $\lambda$=3, $\beta$=0.001, $\gamma$=0.7, $\alpha$=0.8 and $p$=0.995. Sub-Figs.(a)-(e) display the distributions based on the original parameters changing $\alpha$, $\gamma$, $\lambda$, $\beta$ and $p$ orderly.}
\end{figure*}

The queueing system of the epidemic process is decided by five parameters, the input rate $\lambda$, the infected rate $\beta$, the output rate $\alpha$, the recovered rate $\gamma$, and the reviving proportion $p$. In the initial system, only one is an infected individual and others are susceptible. Let time be 4000 which is long enough for the propagation process to be stationary.

The individual number varying with time is shown in Fig. \ref{fig:n123}, where plots of the number of individuals all tend to be stationary, fluctuating around a certain value after some time. In all sub-figures in Fig. \ref{fig:n123}, the blue curve denotes the number of $S$ varying with time, the red curve indicates the number of $I$, and the green one indicates the number of $R$. And we can see that in all these sub-figures, during a very short period from the beginning, the susceptible individuals grow rapidly while the infected and recovered individuals increase slowly. Then, the number of susceptible individuals decrease abruptly while at the same time, the number of individuals in infected and recovered state both increase. After $t$=1500, the number of individuals in three states becomes stationary. Fig. \ref{fig:n123}(a) displays the system with $\lambda$=3, $\beta$=0.001, $\gamma$=0.7, $\alpha$=0.8, and $p$=0.995. The individual number in three states $S$, $I$ and $R$ eventually approaches 200, 750, and 850 approximately.

The system shown in Fig. \ref{fig:n123}(b) is set $\alpha=1.6$ which is twice larger than that in Fig. \ref{fig:n123}(a), other parameters are the same. We can see that the recovered individual number fluctuates around 400 which is about half of the number compared to the system in Fig. \ref{fig:n123}(a). This indicates that a larger $\alpha$ results in less recovered individuals in the system. Fig. \ref{fig:n123}(c) illustrates the system with a twice higher input rate that is $\lambda$=6. The number of infected and recovered individuals respectively fluctuates between 1500 and 1750, which are twice higher than those in Fig. \ref{fig:n123}(a). The number of susceptible individuals in Fig. \ref{fig:n123}(d) is halved and converges to 100 approximately. In Fig. \ref{fig:n123}(e), the stationary number of susceptible individuals gets doubled, surging to above 300 in inverse with the stationary number of $I$ decreasing to above 400, half of that in Fig. \ref{fig:n123}(a). And finally, we can see from the number of infected and recovered individuals in Fig. \ref{fig:n123}(f) both halved compared to that in Fig. \ref{fig:n123}(a), fluctuating around 400 which is above the number of $I$ while below the number of $R$.

We next analyze unchanged curves in Figs. \ref{fig:n123}(a)-(e). The susceptible individual numbers in \ref{fig:n123}(b), (d) and (f) are identical to that in Fig. \ref{fig:n123}(a) which indicates $\alpha$, $\lambda$ is independent of $S$. And the infected individual numbers in Figs. \ref{fig:n123} (b) and (d) are the same as Fig. \ref{fig:n123}(a), indicating that $\alpha$ and $\beta$ have no influence on the number of $I$. The recovered individual number in Figs. \ref{fig:n123}(d) and (e) witness the same stationary value as that in Fig. \ref{fig:n123}(a), which verifies the recovered individual number is independent of $\beta$ and $\gamma$.

For a better illustration, we calculate the average value of individual number of three states when the system is stationary and make a comparison to theoretical results. The weight $w_{n}$ of the individual number $n$ depends on the sum of its remaining time. And we observe the number of individuals from $t_{b}=3000$ to $t_{e}=4000$. We obtain the expected values of records during this time via
\begin{equation}
E(x)=\frac{t_{n}}{t_{b}-t_{e}}x,
\end{equation}
where $t_{n}$ is the time length of the period when the individual number is $n$ and $t_{s}-t_{e}$ is the total time. Besides, the standard deviation of records is also calculated by
\begin{equation}
\frac{1}{N}\sqrt{\sum(x-E(x))^{2}}.
\end{equation}
We also calculate the absolute percentage error for a comparison between theoretical expectation and expected values in simulations. The results are presented in Tabs. \ref{tab:S}, \ref{tab:I} and \ref{tab:R}, which respectively presents the results of $S$, $I$ and $R$. There are three couples of ($\beta$, $\gamma$) in Tab. \ref{tab:S}, four couples of ($\lambda$, $\gamma$, $p$) in Tab. \ref{tab:I} and four couples of ($\lambda$, $\alpha$, $p$) in Tab. \ref{tab:R}. As we can see in the three tables, the simulation results are in accord with theoretical results. The standard deviations in Tab. \ref{tab:S} are less than 0.123, and the standard deviations are less than 0.128 and 0.131 respectively in Tabs. \ref{tab:I} and \ref{tab:R}. The small deviation indicates that the number of individuals in three states is around the stationary value in a small range. And the maximum value of absolute percentage errors is respectively 1.9$\%$, 3.5$\%$ and 3.7$\%$, which demonstrates the accuracy of our theoretical expectation described in Th.\ref{ty2}.



In addition, we illustrate the distribution of the number of individuals in $S$, $I$ and $R$ respectively given by different parameters. Different values are also set for $\alpha$, $\gamma$, $\lambda$, $\beta$ and $p$ displayed in Figs. \ref{fig:ds}, \ref{fig:di} and \ref{fig:dr}, where the results present an obvious normal distribution in form.

In Fig. \ref{fig:ds}, we demonstrate the distribution of individuals of $S$. As we proved in above Th. \ref{ty2}, the expectation of the stationary number of $S$ is independent of $\lambda$, $\alpha$ and the proportion $p$, which results in that they are identical distributed in Figs. \ref{fig:ds}(a), (b) and (c). And in Fig. \ref{fig:ds}(d), the distribution marked with green circles is wider than the blue one, and the distribution marked with red crosses is the widest that indicates the largest variance. Besides, the blue-scatter diagram with the small $\lambda$ is on the left side, whose mode is in the middle of 50 and 100, while the red distribution with the largest $\lambda$ is rightmost whose mode is close to 300. The green scatters with the intermediate value of $\lambda$ is between the other two distributions. This indicates that the larger value of $\gamma$ is, the larger number of the susceptible individual is. Fig. \ref{fig:ds}(e) shows a different relation between the distribution and the parameter $\beta$, compared to ${\gamma}$, which proves that the large value of the infected rate $\beta$, however, leads to the decrease of the susceptible individuals.

The distributions of individuals of $I$ are shown in Fig. \ref{fig:di}. As is illustrated in Figs. \ref{fig:di}(a) and (b), the distributions are mostly overlapped and the number of individuals of peak value is identical, indicating that the individual number of $I$ is independent of the output rate $\alpha$ and the $\beta$. While in Fig. \ref{fig:di}(c) we can clearly see the positions on the horizontal ordinate corresponding to the peak values of the distribution are different. In detail, the expectation value of distribution with $\gamma$=0.35 marked by the blue triangle is between 1500 and 1750, while the distribution with $\gamma$ twice higher than the blue one has a peek-relevant value between 750 and 1000, closer to 750. And the distribution with $\gamma$=3.5 is marked by red crosses, whose value corresponding to the peek is between 500 and 750, closer to 500. This indicates the infected individual number is inverse proportion to $\gamma$. In contrast, from Fig. \ref{fig:di}(d), the individual number values of the peak are the same as that in Fig. \ref{fig:di}(c) corresponding to $\lambda$ that are 6, 3, 1.5, showing a in-proportion relation with $\lambda$. According to Fig. \ref{fig:di}(e), the distribution with the smallest $p$ is leftmost while with largest $p$ is rightmost, verifying that a larger value of $p$ contributes to more infected individuals.

Fig. \ref{fig:dr} displays the distribution of the individual number in $R$. According to Fig. \ref{fig:dr}(a), $\alpha$=1.6 corresponds to the position of the peak on the horizontal ordinate is lower than 400, $\alpha$=0.8 corresponds to lower than 800, and the value on the horizontal ordinate of the peak with $\alpha$=0.4 get doubled approximately, matching the value between 1400 and 1600, which indicates $\alpha$ is inverse proportion to the number of $R$. In addition, the distribution with $\alpha$=0.4 has a bigger width than that with $\alpha$=0.8 and $\alpha$=1.6, whose variance is larger. On the contrary, the distribution with different $\lambda$ and $p$ is shown in Figs. \ref{fig:dr}(b) and (c) where $\lambda$=1.5, 3, 6 corresponds to the value between 250 and 500 closer to 500, the value between 750 to 1500, closer to 750 and value approaching 1750 approximately, and $p$=0.990, 0.995, 0.997 corresponds to 400, 800, 1200, both of which presenting a positive influence on the individual number of $R$. Large values of $\lambda$ and $p$ cause the distribution is on the right of the ordinate, namely, accelerating the number of individuals of $R$. Besides, the distributions in Figs. \ref{fig:dr}(d) and (e) both are overlapping, which indicates that the distribution of the individual number in $R$ is independent of $\gamma$ and $\beta$.

Consequently, the above simulation results agree with the theoretical results as described in Th. \ref{ty2}. Besides, in the light of the individual protection degree that we proposed in the first model, people enhance the awareness of protection and take measures to protect themselves, e.g., wearing masks or avoiding crowds can also suppress the epidemic level. And according to the second model reducing the input rate $\lambda$ and the reviving probability $p$ can make the infected individual number decreasing, regimentation, e.g., strengthening the administration of transient population of an area thus should be encouraged to inhibit the epidemic.

\subsection{Simulation of a second wave of COVID-19 in Zhengzhou}
\begin{figure}[!t]
\includegraphics[scale=0.35]{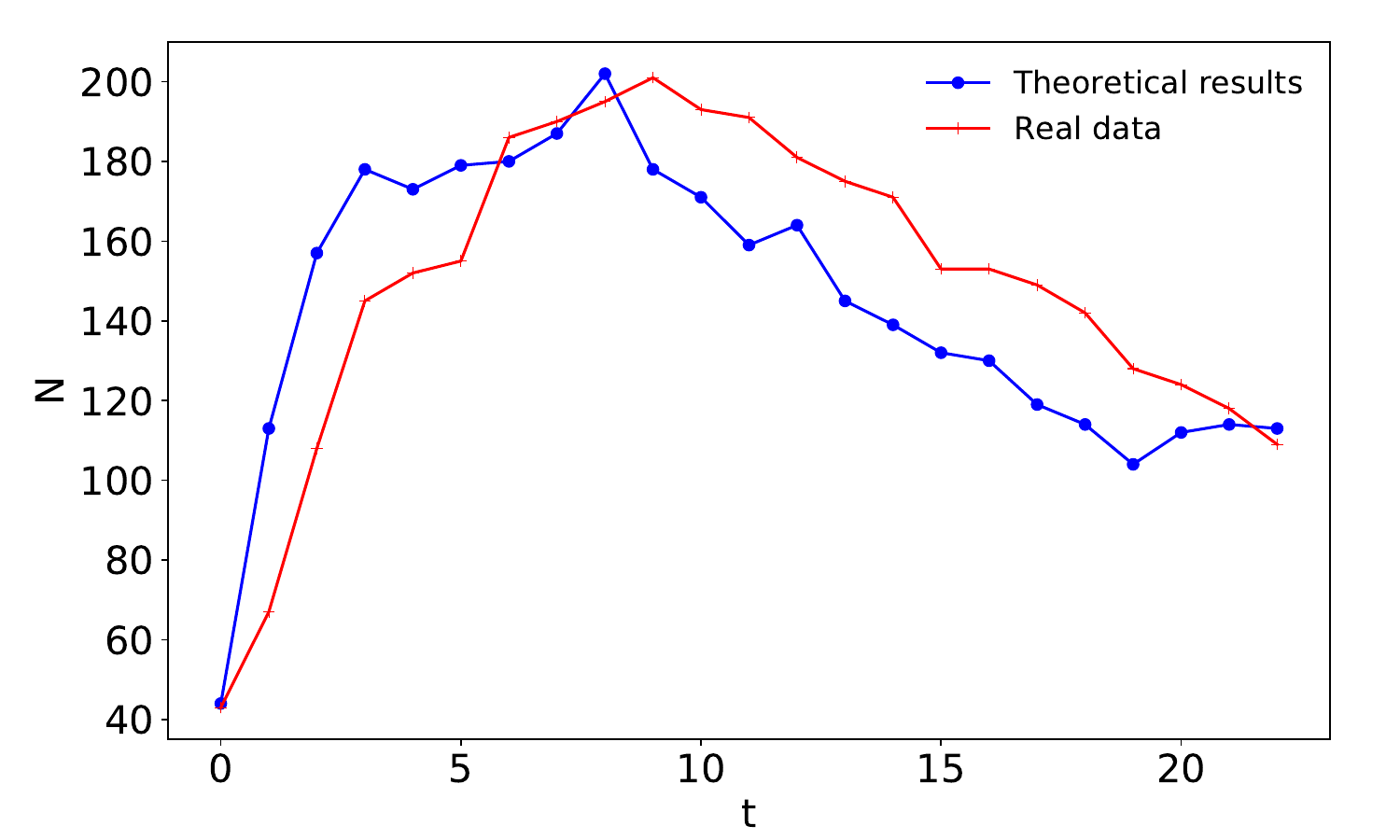}
\centering
\caption{The comparison between real data and theoretical results: The red cross plot is the real infected individual number during the time window, and the blue circle plot is the theoretical infected individual number according to the population model. The two curves have the same trends on the whole.}
\label{fig:real}
\end{figure}

In this subsection, we simulate the number of infected individuals in Zhengzhou, Henan province, China by utilizing the SIRS population model. According to our proposed model based on the queueing system, the infected individual number is the number of individuals in the I service center. In order to control the second wave of epidemics in Henan, the specific areas are under restriction, where people are allowed getting into the area but leaving only for essential trips. Therefore, we simulate the infected individual number in the restricted area which is regarded as an open Markov queueing network in our population model. Based on the data set of COVID-19, the second wave in Henan began on 31st July when there were domestic infected cases appearing. The restriction for the high-risk area began on 3rd August. We take the time window from 6th August to 28th August which lasts 23 days, and mainly observe the epidemics during this time window. We set the $t_0$=0 and the termination $T$=22. It takes time to be diagnosed, which leads to the real data delayed, while our theoretical results are timely, hence, we shift the theoretical results backward in time when comparing them.

We next analyze the real situation and apply our model to real data. Most confirmed cases are in Zhengzhou, thus we take the data of Henan province as that of Zhengzhou. In the real situation of the case in Zhengzhou, most people have been vaccinated, hence, a large fraction of individuals are in the recovered state. However, the immunity got from the vaccine is not absolutely valid for the virus especially for the mutated ones. There is still the probability to get infected once contacting the infected individuals. Therefore, we regard all individuals in the area as susceptible individuals with a quite large protection degree to describe the situation that being vaccinated but not absolutely valid. Additionally, we suppose that the individuals involved are close contacts numbered 2515 which compose the underlying network of the epidemics in the high-risk area, and suppose all these individuals are in the restricted area. For the parameters in the population model, we then set appropriate values for them. In terms of the network setting, the population network is regarded as a small-world network with the total individual number $n$=2000 and the initial nearest neighbor are both one on the left and right side. We suppose that the fraction $p$ of recovered individuals being susceptible again is 0 during the second wave, then $\alpha$ is the output rate excluding the reviving flow. The input and output rate are set to be very small due to the high risk, especially for the output rate is even smaller than the input rate under restriction, and we set $\lambda$=0.01 and $\alpha$=0.005. The protection degree is set to follow a Poisson distribution with the intensity $\mu$=1 which is the expectation. The infected rate $\beta$ becomes $\beta\mu$ seven days after the first individual being infected. The recovered rate $\gamma$ is set to be 0.55.
\begin{table}[t]
  \centering
  \caption{The similarity of real data and theoretical results under three measures.}
    \begin{tabular}{|c|c|c|c|}
    \hline
    Measure & Pearson coefficient & Cosine similarity & CORT \\
    \hline
    Value & 0.800 & 0.987 & 0.656 \\
    \hline
    \end{tabular}%
  \label{tab:realp}%
\end{table}%

The real data and theoretical results are illustrated in Fig. \ref{fig:real}. The red cross plot denotes the infected individual number (daily stored confirmed cases), and the blue circle plot denotes the theoretical infected individual number based on our model. We can see that the two plots have a similar trend.
To further make a comparison between the real data and the theoretical results, we introduce three methods of comparing the similarity of two plots of the infected individual number, which are the Pearson correlation coefficient, the cosine similarity, and the first order temporal correlation coefficient (CORT). The Pearson correlation coefficient is expressed as
\begin{equation}
\rho=\frac{E[(X-E(X))(Y-E(Y))]}{SXSY},
\end{equation}
where $X$ denotes the real data, $Y$ denotes the theoretical results, $E$ is the expectation function
and $S$ is the standard deviation. The value of the coefficient is between -1 and 1, where 1 presents a totally positive linear correlation, -1 indicates a totally negative linear correlation, and 0 indicates no linear correlation.

The cosine similarity is expressed as
\begin{equation}
cos=\frac{\sum_{i=1}^{n}X_{i}Y_{i}}{\sqrt{\sum_{i=1}^{n}X_{i}^{2}}\sqrt{\sum_{i=1}^{n}Y_{i}^{2}}},
\end{equation}
where $X_i$ is a component of the real data vector $X$, $Y_i$ is a component of the theoretical results vector $Y$. The value of the cosine similarity is between -1 and 1, 1 indicates that the vector angle is 0, and -1 indicates that the vector angle is 180.

The CORT is expressed as
\begin{equation}
CORT=\frac{\sum_{t=1}^{T-1}(X_{t+1}-X_{t})(Y_{t+1}-Y_{t})}{\sqrt{\sum_{t=1}^{T-1}(X_{t+1}-X_{t})^{2}}\cdot\sqrt{\sum_{t=1}^{T-1}(Y_{t+1}-Y_{t})^{2}}},
\end{equation}
where $X_{t}$ is the real infected individual number at time $t$, and $Y_{t}$ is the theoretical infected individual number at time $t$. The value of CORT is between -1 and 1, where 1 presents the same trend, -1 indicates the opposite trend, and 0 indicates no temporal correlation.

The real data is recorded by the time unit of one day, which is a discrete sequence. Thus, we also take our theoretical results as a discrete sequence divided by the same time unit. The comparison is demonstrated in Tab. \ref{tab:realp}, from which we can see that the Person coefficient is 0.800, indicating a strong linearly dependent relation between the real data and theoretical results. Besides, the cosine similarity is 0.987 which is quite high. The value of CORT is 0.656 which is not as large as that of the Pearson coefficient and the cosine similarity the reason for which is that our theoretical results varying with fluctuation, while it is yet over 0.5. These verify that the real data and the theoretical results agree with each other.


\section{\label{sec:IV}Conclusions and Outlook}
In this paper, we study the SIRS epidemic model from two perspectives based on statistical methods. In the first model, in terms of individual, we introduce the individual protection degree considering the protective awareness in an epidemic and obtain the numerical value of stationary probability that an individual is in $S$, $I$, and $R$ state. In the second model regarding the individual number, we consider the migration of mobile people. A Markov queueing system is constructed to model the migration of people and transitions between epidemic states. Distinct from the number of individuals converging to a constant value obtained by primer studies, we achieve that the individual number in each epidemic state is convergent in probability by demonstrating the limited distributions by simulations and also obtain the expectation value. Moreover, the protection degree suppresses the epidemic level according to our model and the impact of parameters in our model, e.g., the input rate of a place and the recovered rate of infected people have a strong influence on the individual number, which inspires us to provide some suggestions for epidemic controlling theoretically.

Nevertheless, there are still some issues to be further addressed. For heterogeneous networks, it is complicated to give the concrete expression to generalize the increments of the number of individuals varying with time. Analytical solutions for the limited probability of nonhomogeneous Markov Chain have been a tough problem all the time. Besides, for practical applications, there is a requirement to apply our population model to fit realistic data. These issues will be further studied in our future work.


\begin{IEEEbiography}[{\includegraphics[width=1in,clip,keepaspectratio]{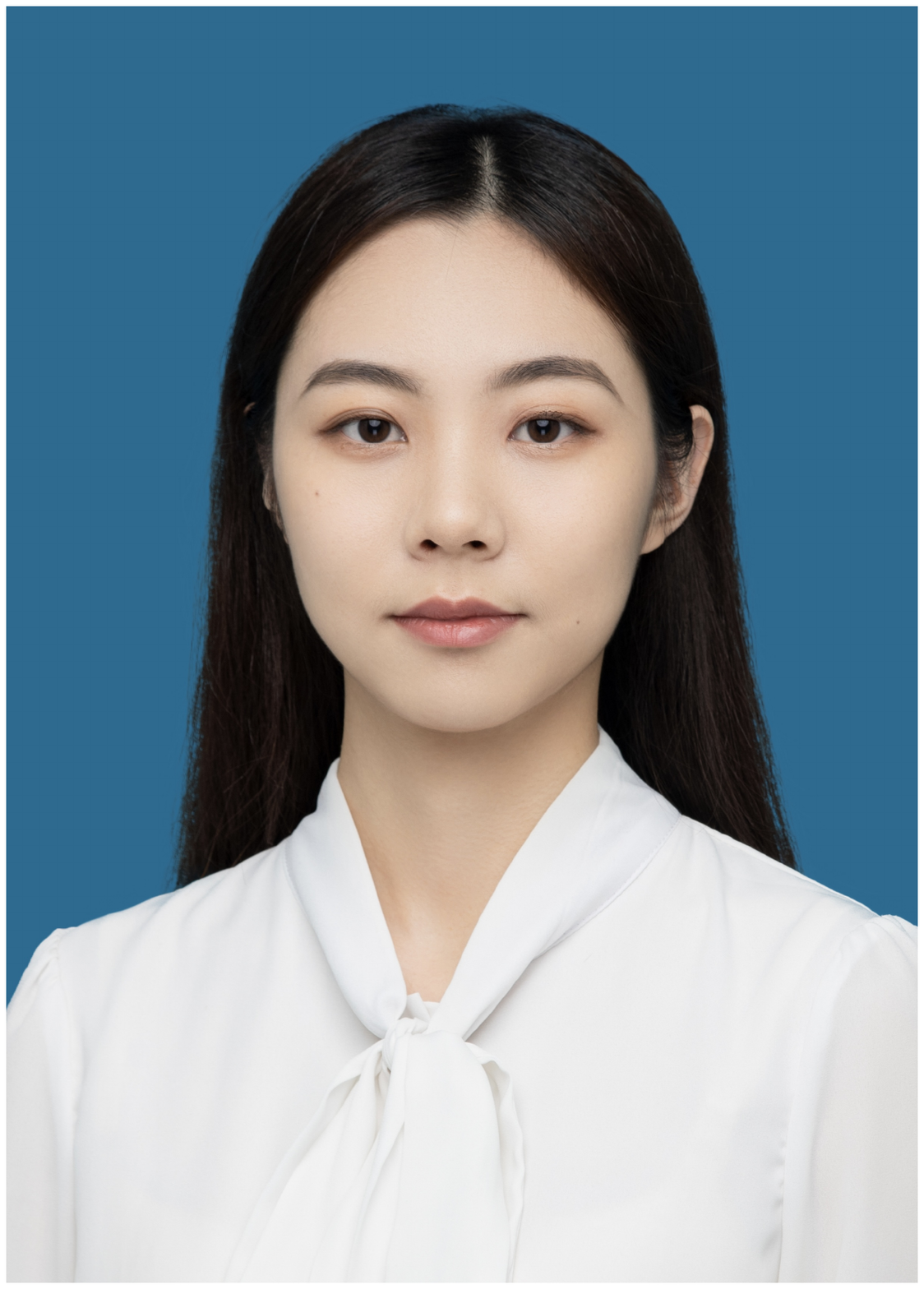}}]{Yuhan Li}
is an undergraduate from the College of Artificial Intelligence, Southwest University, Chongqing, China. Her research interests include stochastic process, complex networks, network propagation and nonlinear science.
\end{IEEEbiography}

\begin{IEEEbiography}[{\includegraphics[width=1in,clip,keepaspectratio]{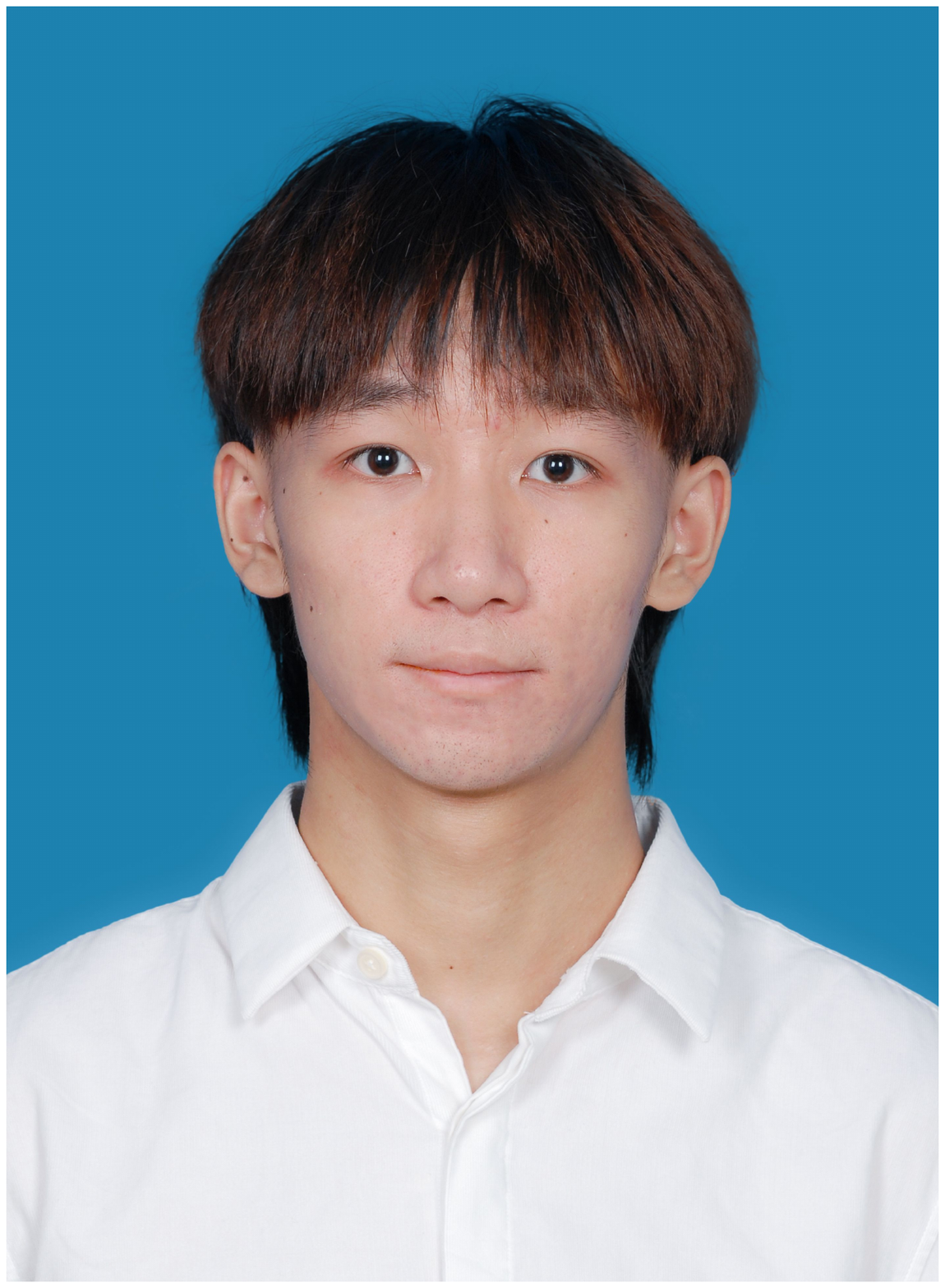}}]{Ziyan Zeng}
is an undergraduate from the College of Artificial Intelligence, Southwest University, Chongqing, China. His research interests include complex networks and evolutionary games and nonlinear science.
\end{IEEEbiography}

\begin{IEEEbiography}[{\includegraphics[width=1in,clip,keepaspectratio]{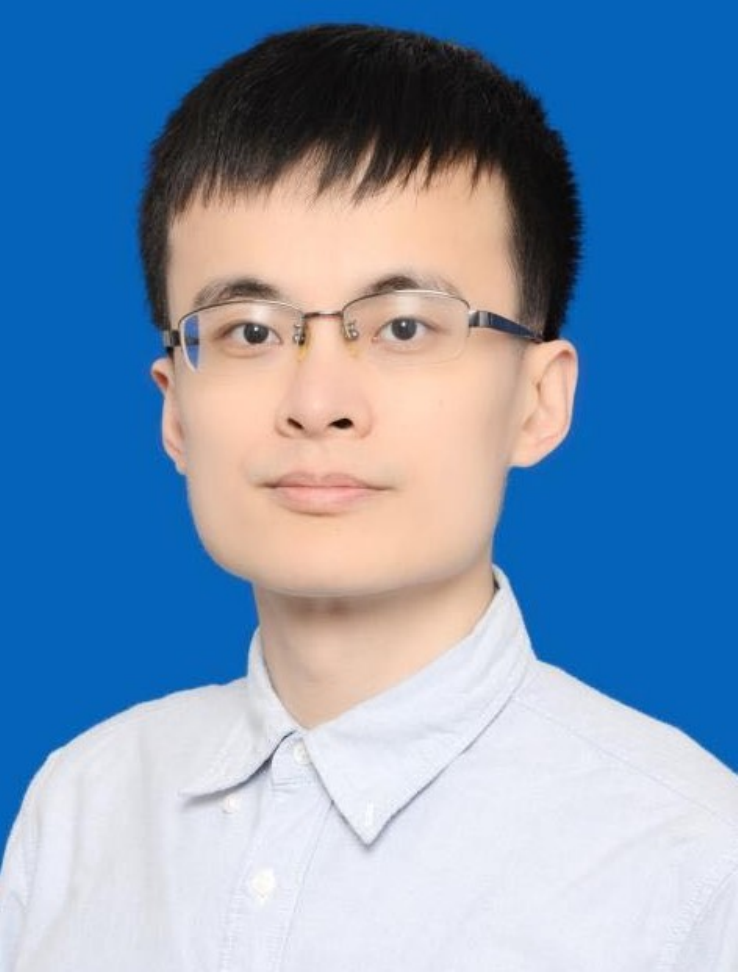}}]{Minyu Feng}
received B.S. degree in mathematics from the University of Electronic
Science and Technology of China in 2010; the Ph.D. degree in computer science from the University of Electronic
Science and Technology of China in 2018. Since 2019, he has been an associate professor in the College
of Artificial Intelligence, Southwest University, Chongqing, China. His research interests include stochastic process, complex networks, nonlinear science and social computing.
\end{IEEEbiography}

\begin{IEEEbiography}[{\includegraphics[width=1in,clip,keepaspectratio]{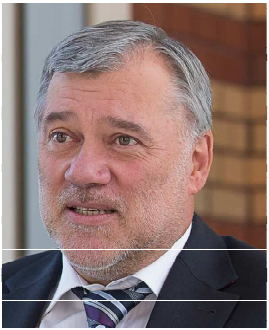}}]{J\"{u}rgen Kurths}
received the B.S. degree in mathematics from the University of Rostock; the Ph.D. degree from the Academy of Sciences of the German Democratic Republic in 1983; the Honorary degree from N.I. Lobachevsky State University of Nizhny Novgorod in 2008; and the Honorary degree from Saratow State University in 2012.

From 1994 to 2008, he was a Full Professor
with the University of Potsdam, Potsdam, Germany.
Since 2008, he has been a Professor of nonlinear
dynamics with Humboldt University, and the Chair
of the Research Domain Complexity Science with the Potsdam Institute
for Climate Impact Research, Germany. He is the author of
more than 700 papers, which are cited more than 53.000 times (H-index: 106).
His main research interests include synchronization, complex networks, time
series analysis, and their applications.

Dr. Kurths is a Fellow of the American Physical Society, of the Royal Society of Edinburgh and of the Network Science Society and a Member of
the Academia Europaea. He was the recipient of the Alexander von Humboldt
Research Award from India, in 2005, and from Poland in 2021, the Richardson medal of the European
Geophysical Union in 2013, and eight Honorary Doctorates. He is a highly
cited Researcher in Engineering. He is editor-in-chief of CHAOS and on
the editorial boards of more than 10 journals.


\end{IEEEbiography}

\end{document}